\documentclass[sigconf,authorversion]{acmart}
\usepackage{acmart-taps}

\usepackage{graphicx}
\usepackage{longtable}
\usepackage{textcomp}
\usepackage{balance}
\usepackage{xcolor, colortbl}
\usepackage{comment}
\usepackage{enumitem}
\usepackage{xspace}
\usepackage{arydshln}
\usepackage{multirow}
\usepackage{booktabs}
\usepackage{fontawesome}
\usepackage{pifont}
\let\oldding\ding%
\renewcommand{\ding}[2][1]{\scalebox{#1}{\oldding{#2}}}%

\definecolor{green-include}{RGB}{49,163,84}
\definecolor{red-exclude}{RGB}{203,24,29}

\usepackage{tikz}
\usetikzlibrary{shapes,arrows}
\usetikzlibrary{positioning,trees
,shapes.geometric, arrows, calc, fit, shadows, intersections,tikzmark}
\usepackage{balance}

\usepackage{backref}

\newcommand{\quoting}[1]{``\emph{#1}''}

\newcommand{\sd}{\emph{SD}=}

\newcommand{\psmall}{\emph{p}$<$0.001}

\newcounter{mycounter}
\newcommand{\inc}{%
{%
\themycounter%
\addtocounter{mycounter}{1}%
}%
}
\newcommand{\increment}{\#\inc}

\definecolor{people}{HTML}{abdda4}
\newcommand{\grouppeople}{\textcolor{people}{\ding[1.1]{108}}}

\definecolor{two}{HTML}{c7e9b4}

\definecolor{tech}{HTML}{3288bd}
\newcommand{\grouptech}{\textcolor{tech}{\ding[1.1]{108}}}

\definecolor{context}{HTML}{f46d43}
\newcommand{\groupcontext}{\textcolor{context}{\ding[1.1]{108}}}

\definecolor{activities}{HTML}{ffeda0}
\newcommand{\groupactivities}{\textcolor{activities}{\ding[1.1]{108}}}

\definecolor{goals}{rgb}{1.0, 0.33, 0.64}
\newcommand{\groupgoals}{\textcolor{goals}{\ding[1.1]{108}}}

\newcommand{\rev}[1]{{#1}}

\AtBeginDocument{%
  \providecommand\BibTeX{{%
    \normalfont B\kern-0.5em{\scshape i\kern-0.25em b}\kern-0.8em\TeX}}}

\copyrightyear{2024}
\acmYear{2024}
\setcopyright{acmlicensed}\acmConference[CHI '24]{Proceedings of the CHI Conference on Human Factors in Computing Systems}{May 11--16, 2024}{Honolulu, HI, USA}
\acmBooktitle{Proceedings of the CHI Conference on Human Factors in Computing Systems (CHI '24), May 11--16, 2024, Honolulu, HI, USA}
\acmDOI{10.1145/3613904.3642124}
\acmISBN{979-8-4007-0330-0/24/05}

\begin{document}

\title{Born to Run, Programmed to Play: Mapping the Extended Reality Exergames Landscape}

\author{Sukran Karaosmanoglu}
\orcid{0000-0002-9624-4258}
\affiliation{
  \institution{Human-Computer Interaction, Universität Hamburg}
  \city{Hamburg}
  \country{Germany}
}
\email{sukran.karaosmanoglu@uni-hamburg.de}

\author{Sebastian Cmentowski}
\orcid{0000-0003-4555-6187}
\affiliation{
\institution{HCI Games Group, Stratford School of Interaction Design and Business, University of Waterloo} 
\city{Waterloo}
\country{Canada}}
\email{sebastian.cmentowski@uwaterloo.ca}

\author{Lennart E. Nacke}
\orcid{0000-0003-4290-8829}
\affiliation{
\institution{HCI Games Group, Stratford School of Interaction Design and Business, University of Waterloo} 
\city{Waterloo}
\country{Canada}}
\email{lennart.nacke@acm.org}

\author{Frank Steinicke}
 \orcid{0000-0001-9879-7414}
\affiliation{
  \institution{Human-Computer Interaction, Universität Hamburg}
 \city{Hamburg}
  \country{Germany}
 }
 \email{frank.steinicke@uni-hamburg.de}

\renewcommand{\shortauthors}{Karaosmanoglu, et al.}

\begin{abstract}
Many people struggle to exercise regularly, raising the risk of serious health-related issues. Extended reality (XR) exergames address these hurdles by combining physical exercises with enjoyable, immersive gameplay. While a growing body of research explores XR exergames, no previous review has structured this rapidly expanding research landscape. We conducted a scoping review of the current state of XR exergame research to (i) provide a structured overview, (ii) highlight trends, and (iii) uncover knowledge gaps. After identifying 1318 papers in human-computer interaction and medical databases, we ultimately included 186 papers in our analysis. We provide a quantitative and qualitative summary of XR exergame research, showing current trends and potential future considerations. Finally, we provide a taxonomy of XR exergames to help future design and methodological investigation and reporting.

\end{abstract}



\maketitle

\section{Introduction}
Technological advances have changed our lifestyle in an unprecedented way. Today, many people spend most of their day in front of a computer, working in cognitively demanding but physically underwhelming jobs. According to the World Health Organization (WHO), 27.5\% of adults and even 81\% of adolescents do not meet the recommended physical activity levels~\cite{WHO2022_globalstatus}. This sedentary lifestyle affects our well-being: insufficient physical activity is associated with severe health issues, such as cardiovascular diseases, dementia, or depression~\cite{WHO2022_globalstatus}. Consequently, the WHO launched the Global Action Plan on Physical Activity (GAPPA) to raise awareness of the importance of regular exercise~\cite{WHO2018_gappa}. Unfortunately, establishing and maintaining a healthy, physically active lifestyle is challenging. It requires time and motivation to become a habit~\cite{pate2011overcoming}. However, modern technology is not only the problem's cause but can also contribute to its resolution. A promising way of supporting people's efforts towards a healthier lifestyle is to use digital games: exergames. These games combine enjoyable gameplay with physical activities to achieve engaging experiences. Although this genre is not in its infancy anymore and includes several milestone games, such as Dance Dance Revolution~\cite{dancedancerevolution} or Wii Fit~\cite{wiifit}, recent technical innovations have inspired a new wave of research on exergames. 

The key to the next generation of exergames is extended reality (XR). XR is an umbrella term for virtual reality (VR), augmented reality (AR), augmented virtuality (AV), and mixed reality (MR)~\cite{ratcliffe2021extended,plopski2022eye,hirzle2023when}. These technologies offer invaluable benefits to designing engaging and responsive exergames for full-body exertion. Above all, the unique advantage of XR is its spatial nature. Physical interactions in a three-dimensional (3D) environment are the foundation of most XR applications. Exergame developers can extend these interaction patterns to blend exercises naturally into the gameplay. Unlike previous technologies, modern XR systems feature built-in six degrees of freedom (6-DOF) tracking, which is often essential for exergames. Furthermore, view-dependent, stereoscopic images improve the perceived realism and immersion in the virtual environment, thereby increasing motivation in VR exergames~\cite{493}. XR technologies also grant full control over the players' surroundings, which can be used to create highly engaging scenarios that are impossible in the real world (e.g., superhuman powers~\cite{8}). \rev{Lastly, the full control over virtual worlds is ideal for customization and personalization: XR exergames can be tailored to meet users' exercise needs (e.g., to design exercise scenarios for people with dementia~\cite{12}). These advantages make XR an ideal fit to take fitness applications to the next level. At the same time, the apparent differences compared to more traditional exergames (e.g., free 360\textdegree view, limited proprioception, focus on motion-based interactions) create significant challenges and open questions that need addressing to ensure the games' safety, efficiency, and enjoyability.}

\rev{The release of more broadly adopted, consumer-ready hardware,} such as Meta Quest 2~\cite{oculusquest2} or Microsoft HoloLens 2~\cite{hololens}, has shifted focus to this promising domain. As a result, the field is producing plenty of promising new contributions. \rev{Despite the breadth of research, a comprehensive, systematic organization of the field is currently lacking.} This creates a complex landscape; it is difficult to get a complete picture \rev{of the areas already covered and the open questions.} \rev{This lack of structure hinders the information flow between different domains in this interdisciplinary field (e.g., movement and computer science). Further, it makes it difficult to identify research trends and promising directions and prevents researchers from following a systematic and efficient approach to advance our knowledge. Providing an organization of such a rapidly growing area is crucial and timely to help the field progress. By conducting a scoping review, we provide practitioners with invaluable resources for designing XR exergames. Most importantly, our results help researchers, especially those early in their careers, to think critically and identify promising research directions.} %

While previous research covered specific aspects of exergames, no prior review or taxonomy has been established for organizing the growing field of XR exergames. \rev{Existing frameworks of non-immersive exergames are not easily applicable to the XR domain due to the significant differences in design. While XR exergame designers can profit from increased immersion and enjoyment~\cite{493}, they must consider unique challenges, such as safety concerns or tracking problems~\cite{41}. Prior reviews also do not include the latest advances in this rapidly evolving field (over 50\% of our identified corpus was published since 2021). Accordingly,} our goal is to create a taxonomy and analysis of XR exergames research using an adapted version of the foundational People-Activities-Contexts-Technologies (PACT) principles~\cite{benyon2005designing} of the human-computer interaction (HCI) field. With our work, we approach these concerns and contribute a\rev{n up-to-date,} comprehensive analysis of the existing body of XR exergame research. To answer our research questions, we conducted a scoping review and identified 1318 relevant papers. Following an eligibility step, we quantitatively (i.e., by frequency reporting) and qualitatively analyzed the final corpus of 186 papers. Focusing on Goals, People, Exercises, Design, and Technologies (GPEDT) principles (inspired by and adapted from PACT~\cite{benyon2005designing}), we created a hierarchical taxonomy of XR exergames following the taxonomy development method of~\citet{nickerson2013method}. 

With our taxonomy, we disassemble the current landscape of XR exergames into a complete set of dimensions. \rev{This taxonomy illustrates the primary goals and design elements explored by domain researchers and reveals trends and underresearched areas. We use these insights to derive} \rev{nine central} research directions that guide the design, implementation, and future study of XR exergames. \rev{Together, these contributions help researchers and practitioners gain an overview of the field, identify promising research questions, and establish interdisciplinary collaborations.}
\rev{Further, our analysis of the XR exergame corpus also highlighted another key challenge for future research: clear reporting of exergame design and study findings. We identified significant differences in terminology and reported information impeding comprehensibility, reproducibility, and transferability. To address this issue, we provide \rev{guiding questions} that help in the systematic reporting of research on XR exergames. We believe that establishing reporting standards will contribute to clear communication in this young and emerging field.} We hope to facilitate the compilation of existing knowledge and spark new research initiatives that will advance the capabilities of XR systems as a platform for gamified exercise. To summarize, our contributions are as follows:
\begin{itemize}[topsep=0pt, noitemsep, leftmargin=+.2in]
    \item a comprehensive overview of the XR exergame research,
    \item a taxonomy of XR exergames based on \groupgoals~Goals, \grouppeople~People, \groupactivities~Exercises, \groupcontext~Design, and \grouptech \xspace~Technologies meta-characteristics,
    \item \rev{nine} future research directions for the field, and
    \item a set of \rev{guiding questions} to formalize the reporting of future XR exergame research.
\end{itemize}

\section{Background}
This section provides an overview of exergames and XR, and explains how authors in previous literature combined those domains to create playful exercise applications.

\subsection{Exergames}
Research uses synonymous terms to describe applications that involve movement-related gameplay, such as exergames~\cite{12}, exertion games~\cite{muller2011designing}, or motion-based games~\cite{gerling2012full}. In this paper, we use the term \quoting{exergames} to refer to \quoting{digital game[s] where the outcome [...] is predominantly determined by physical effort}~\cite{muller2011designing}. In our review, we consider any game that matches this specification \rev{and is defined} as an exergame \rev{(or variant words)}, regardless of its physical activity level (e.g., breathing, high-intensity interval training). 

Since the \rev{1990s}, exergames have been among the top games. In Dance Dance Revolution~\cite{dancedancerevolution}, players perform dance-inspired rhythmic stepping movements. Another notable example was the Wii console bringing various exergames, called WiiFit games~\cite{wiifit}, to people's living rooms. Later, Microsoft Kinect~\cite{kinect} introduced full-body motion-tracking as a novel input for commercial exergames. This fascinating coupling of digital games with exercising is moving to a new level with novel, immersive XR technologies.

\subsection{Immersive Extended Reality}
 \label{sec:immersiveextendedreality}

The reality-virtuality continuum~\cite{milgram1994taxonomy,milgram1995augmented} is a popular taxonomy for XR-related term definitions. This continuum encompasses the physical real environment on one end and the digital virtual environment on the other. The real environment consists of only real-world objects, whereas the virtual environment (i.e., VR) includes only artificial objects. According to \citet{milgram1994taxonomy}, MR describes the blending of these two realities. \rev{However, according to \citet{speicher2019what}, experts differ in how they understand the term MR, with some using it following the explanation of \cite{milgram1994taxonomy,milgram1995augmented}, while others using different meanings (e.g., strong AR).} We \rev{follow the definition of~\cite{milgram1994taxonomy,milgram1995augmented} and use the term MR} as everything between reality and VR, including AR and AV. AR overlays virtual objects on real surroundings, while AV overlays real objects on virtual environments. \rev{To encompass all these concepts, we use the umbrella term XR. We note that other sources~\cite{rauschnabel2022what} do not understand XR as extended reality but as xReality, with X serving as a placeholder to denote different realities. However, the overall use as an umbrella term remains the same~\cite{ratcliffe2021extended,plopski2022eye,hirzle2023when}.}

Unlike its colloquial use, XR nomenclature solely indicates the relationship between virtual and real content; it does not make assumptions about perceptual effects. To describe how these systems (partially) replace users' sensations with an artificial environment, researchers use the terms \emph{immersion} and \emph{presence}. 
\citet{slater2016enhancing} define immersion as the technical quality of a setup. It is a continuum that depends on various characteristics such as stereoscopic vision, field-of-view, resolution, latency, or sensor substitution. %
These factors allow us to compare XR systems with one another. \citet{slater2016enhancing} suggest that head-mounted displays (HMDs) are more immersive than CAVE systems because they can simulate the latter, but not vice versa. In general, an immersive system should at least provide a head-view-point-dependent, stereoscopic image of the virtual environment~\cite{slater2016enhancing, furht2008immersive}. %
For our review, we follow these definitions and consider any system with stereoscopic vision and an effective display area that exceeds the players' field of view without requiring input other than head rotation. These requirements can be achieved with spatially-tracked HMDs (i.e., VR HMD, AR HMD) or by surrounding the player with projection walls or displays (i.e., CAVE). However, single monitors, floor projections, or hand-held AR do not meet these criteria; \rev{thus we do not consider these to be immersive XR.} 

\subsection{Immersive Extended Reality Exergames}

A growing body of research is exploring the convergence of XR and exergames, highlighting the advantages of XR exergames. \citet{493} showed that VR exergaming provides higher motivation, embodiment, and performance than non-immersive exergaming. Similarly, \citet{367} found that VR exergames improve player performance compared to playing the identical game on a large display. HMD-VR and CAVE-based VR exergaming increase flow and presence compared to non-VR exercise~\cite{579}.

Immersive XR motivates, engages, and enables players to do activities. \citet{8} showed that virtual augmentation of running and jumping contributes to intrinsic motivation. Similarly, \citet{40} found that players perform longer voluntary strenuous activities if they experience augmentation. Further, VR's realism can also be used to build cognitively and physically stimulating exergames for people with dementia~\cite{12}.

With new technological developments---HoloLens 2 released in 2016~\cite{hololens}---immersive AR has begun to be used for exergaming~\cite{265,167,217}. According to a recent study, AR exergaming can lead to a significantly lower level of collision anxiety (i.e., being aware of surroundings) compared to its VR counterpart. This unique AR feature---blending virtual and real environments---has also increased exergame advancements in rehabilitation~\cite{1,100,265}.

\subsection{Existing Literature Reviews \& Taxonomies}

Increasing interest in XR-based exercising has led to more recent review articles. \rev{While some of these works focus on specific subtypes or use cases of XR exergames, no work has yet provided an up-to-date overview of this rapidly evolving domain.}
\rev{Some reviews focused on general physical activity, not particularly on exergames (e.g.,~\cite{odenigbo2022virtual,giakoni2023physical}). Other reviews covered non-immersive ``VR'' exergames (e.g.,~\cite{drazich2020exergames, corregidor2020can,chen2023vr}) and examined their health-related outcomes (e.g., ~\cite{mo2023effects,1029T_lin2023effectiveness,drazich2020exergames}). For instance, \citet{mo2023effects} concluded that exergames are overall safe but not significantly effective for musculoskeletal pain in older adults. \citet{kappen2019older} identified various focuses of non-immersive exergames (e.g., cognitive training) for older adults, which we incorporated in the goal dimension of our taxonomy.  
Lastly, only a few papers considered the intersection of XR technology and physical activity (e.g.,~\cite{giakoni2023physical,odenigbo2022virtual}). \citet{odenigbo2022virtual} reviewed 39 VR, AR, and MR physical activity interventions and showed that most of them included exergaming. Only one paper~\cite{tao2021immersive} provided a narrative review on the overlapping space of XR and exergames, featuring 29 HMD-VR \quoting{health games}. The authors found that most games used obstacle-based gameplay and extrinsic rewards. However, this review focused on health-centered exergames and HMD-based VR, without covering the broader XR field or providing a taxonomy of XR exergames.} 

\rev{Structuring research on sports systems is an ongoing effort in HCI research. 
\citet{reilly2009general} classified computer-augmented sports systems based on two high-level dimensions: form and function. While the function dimension contains the system's purposes and abilities (e.g., sports entertainment, refereeing), the form dimension concerns its implementation (e.g., hardware, software). 
Similarly, \citet{frevel2020taxonomy} present a \emph{SportsTech Matrix}, considering sports and technology from two angles: The user angle captures user groups interacting with sports technology (e.g., athletes, consumers), whereas the tech angle comprises technology used with/for sports.
Inspired by \citet{reilly2009general}, \citet{postma2022design} provide a taxonomy of sports interaction technology to bridge sports science and HCI. For instance, they introduce new forms of sports interaction technology relating to space, time, game nature, feedback, and integration of interaction.
Although these taxonomies hold importance for advancing the field, they do not particularly focus on exergames. Additionally, not all exergames match the characteristics of sports according to \citet{jenny2017virtually}'s definition; for example, an exergame does not need to \quoting{include competition (outcome of a winner and loser)}. Similarly, a sport can be supported by technology without requiring additional gameplay.}

\section{Research Focus: What We Add to the Literature} 
\label{sec:researchfocus}

\rev{Likely most related to our research is the work by \citet{mueller2008taxonomy}, who approach the field through a social lens and present a taxonomy with four dimensions: non-exertion vs. exertion, non-competitive vs. competitive, non-parallel vs. parallel, and combat vs. object. For example, the exertion dimension provides an understanding of what an exergame is (in line with the definition we followed in~\cite{muller2011designing}), whereas the competitive unit covers exergames featuring opponents. While this taxonomy helps to define social aspects in exergames, it does not provide classifications on other aspects and does not target XR technology. However, XR exergames have unique advantages (e.g., immersion, real-time feedback) and challenges (e.g., safety, technical constraints)~\cite{41}.} 
The unique potential of such applications---offering engaging and motivating exercises---and continued advances in XR hardware motivate additional concentrated research. We believe a structured review identifying trends and knowledge gaps could strengthen community efforts.

\rev{We comprehensively review exergames that use XR technology (e.g., CAVE-VR, AR) using a scoping review, and also include non-health exergames to provide an extensive picture of the domain through an HCI lens. With our review, we first provide a summary of XR exergame research (e.g., studies). Based on our review, we then give a quantitative and qualitative analysis of the XR exergames featured in research and create a hierarchical taxonomy. We base this step on the established PACT framework~\cite{benyon2005designing}. This framework serves as a guide for creating interactive systems using a human-centered design (HCD) perspective~\cite{iso}. Since XR exergames are highly interactive and humans are at the core of these systems, we used the PACT framework elements as an inspiration. However, we adapted them to fit the specific characteristics of exergames. We split the Activities dimension into Goals and Exercises since exergames typically have an overarching interventional goal (e.g., rehabilitation) and specific exercises that contribute to this purpose\footnote{\rev{We note that some may see exercises as part of the design. However, since exercises are the critical components of exergames and correspond to activities within the PACT framework, we keep this aspect separate.}}. Also, we renamed Contexts into Design to better reflect which aspect of the exergame we aimed to capture. In total, our GPEDT consists of the five principles: \groupgoals~Goals, \grouppeople~People, \groupactivities~Exercises, \groupcontext~Design, and \grouptech~Technologies. Overall, we summarize our research questions motivating the taxonomy as follows:}

\begin{itemize}[topsep=0pt, noitemsep]
\item[\textbf{(RQ1)}] \groupgoals~\textbf{Goals}: What are the goals of the XR exergames? %
\item[\textbf{(RQ2)}]\grouppeople~\textbf{People}: Which different user groups are usually targeted by XR exergames? 
\item[\textbf{(RQ3)}] \groupactivities~\textbf{Exercises}: What kinds of exercises are being designed in the XR exergames?
\item[\textbf{(RQ4)}] \groupcontext~\textbf{Design}: What kinds of game design aspects are considered for XR exergames?
\item[\textbf{(RQ5)}] \grouptech~\textbf{Technologies}: What kinds of technologies are being used in the XR exergames?
\end{itemize}

\section{Scoping Review of Extended Reality Exergames}
\begin{figure}
{
\Huge
\resizebox{0.5\textwidth}{!}{
\includegraphics{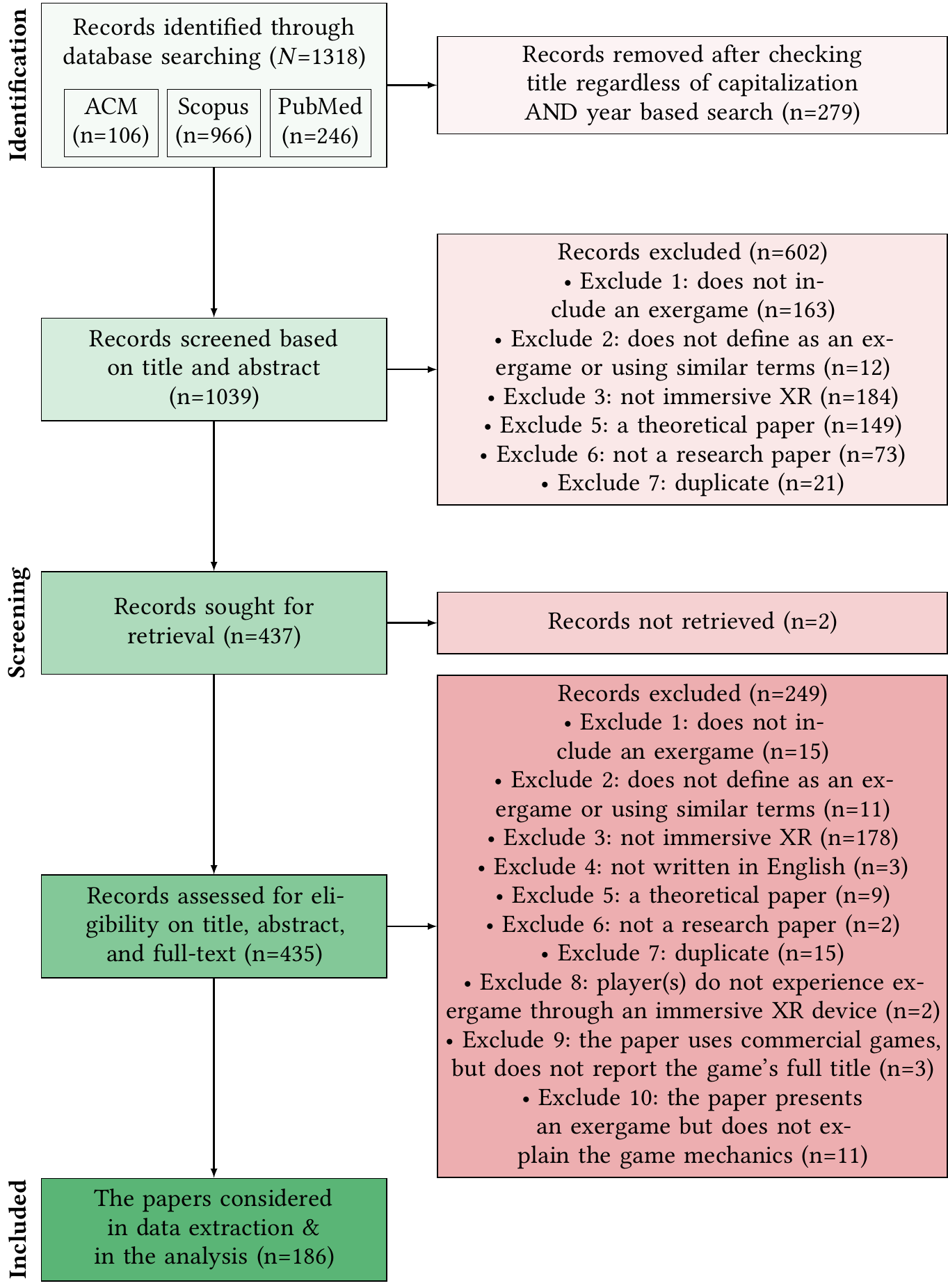}
}}
\caption{We illustrate our scoping review process in a PRISMA flow diagram to detail our steps~\cite{prisma2020flowchart}.}
\label{fig:prisma}
\Description{This figure illustrates the review process in a PRISMA flow diagram. It includes identification, screening, and inclusion steps, resulting in 186 publications that we considered for analysis and data extraction.}
\end{figure}

To assess current XR exergame contributions, we conducted a scoping review~\cite{munn2018systematic, sutton2019meeting}. Scoping reviews provide an overview of a specific problem and identify potential research directions~\cite{munn2018systematic}. These reviews are precursors to systematic reviews and do not involve a critical appraisal stage (typical for systematic reviews) to evaluate the methodological quality of articles~\cite{munn2018systematic}.
For our review, we follow the recommendations of the PRISMA extension for scoping reviews (PRISMA-ScR)~\cite{tricco2018prisma} and best practices of \citet{peters2015guidance,peters2022best}.
Our main steps for the scoping review are (i) identification of the corpus, (ii) screening, (iii) eligibility, (iv) data extraction, (v) data synthesis, and (vi) reporting (see \autoref{fig:prisma}).

\begin{table*}[ht]
\small
    \caption{This table shows the XR- and Exergames-related keywords. Within the same group of keywords, we used the \texttt{OR} operator, while for between the keywords groups, we used \texttt{AND} operator. The final search was conducted on 11 August 2023 in ACM (The ACM Guide to Computing Literature), Scopus, and PubMed digital libraries.}
    \Description{This table shows the XR- and Exergames-related keywords. Within the same group of keywords, we used the OR operator, while for between the keywords groups, we used AND operator. The final search was conducted on 11 August 2023 in ACM (The ACM Guide to Computing Literature), Scopus, and PubMed digital libraries.}
\resizebox{\textwidth}{!}{
    \centering
    \begin{tabular}{p{0.4\textwidth}p{0.1\textwidth}p{0.45\textwidth}}
        \toprule
         \textbf{XR-related keywords}  & &  \textbf{Exergames-related keywords} \\
         \midrule
         \texttt{("immersive" OR "VR" OR "AR" OR "AV" OR "MR" OR "XR" OR "virtual realit*" OR "augmented realit*" OR "extended realit*" OR "mixed realit*" OR "augmented virtualit*" OR "virtual environment*")}
         & \centering\texttt{AND} &
         \texttt{("exergame*" OR "exercise game*" OR "physical game*" OR "movement game*" OR "motion game*" OR "motion-based game*" OR "movement-based game*" OR "training game*" OR "exertion game*" OR "sport game*" OR "sports game*")} \\
         \bottomrule
    \end{tabular}}

    \label{tab:keywords}
\end{table*}

\subsection{Protocol, Databases, \& Search}
Following the best practices of conducting a scoping review~\cite{tricco2018prisma,peters2022best}, we prepared a protocol that reports our steps (see supplementary materials). 
We decided on Association for Computing Machinery (ACM)\footnote{\url{https://dl.acm.org/}}, Scopus (indexing IEEE Xplore and many other HCI-related sources)\footnote{see the Scopus indexed venues: \url{https://www.scopus.com/sources}}, and PubMed\footnote{\url{https://pubmed.ncbi.nlm.nih.gov/}} as our databases because XR exergames is an interdisciplinary research area and mostly HCI and medical science scholars publish in our selected databases. 

We performed informal searches on these databases to uncover exergame definition phrases and include less prevalent synonyms like motion-based games or movement-based games. We combined all identified phrases in our search query using \texttt{OR} operator\rev{s}. XR-related keywords were decided based on \citet{milgram1995augmented}'s reality-virtuality continuum. We also used related phrases like immersive or virtual environments. Similar to the exergames-related keywords, we concatenated all keywords using \texttt{OR} operator\rev{s}. Consequently, our search query contained XR-related keywords \texttt{AND} exergames-related keywords (see \autoref{tab:keywords}). We repeated our search query in the selected databases on different days to \rev{double-check for inconsistencies within the databases~\cite{rogers2021best, macarthur2021you}}. We provide our exact search queries for each database in the Appendix (see \autoref{tab:newquery}); the final search was conducted on August 11, 2023.

During data collection, we considered reviewing commercial games as well. However, we deliberately decided against this step because of XR's spatial interaction nature: Many VR games require at least a little physical effort to complete tasks. Still, this does not mean that they are designed as exergames~\cite{90}. Thus, we let XR exergame researchers decide: if a commercial game is considered an exergame in any research paper, we included it in our corpus.

\subsection{Identified Corpus \& Duplicate Removal}
Our search query had no constraints (e.g., publication date) because we intended to cover the current state of XR exergames research~\cite{munn2018systematic}. Our search yielded 1318 papers (The ACM Guide to Computing Literature = 106, Scopus = 966, PubMed = 246). After the identification of articles, we removed duplicates testing for matching year and title (regardless of capitalization) using Python scripts ($n$=279), leaving a total of 1039 articles for the screening phase. Please refer to our supplementary materials for all papers.

\subsection{Screening}
For the screening step, we first decided on inclusion criteria (IC):
\begin{enumerate}[topsep=0pt, noitemsep, leftmargin=+.2in]
    \item IC1: The paper includes an ``exergame'', ``exercise game'', ``exertion game'', ``physical game'', ``movement game'', ``motion game'', ``motion-based game'', ``movement-based game'', ``training game'', or ``sport(s) game'' application.
    \item IC2: The paper defines the application as an ``exergame'', ``exercise game'', ``exertion game'', ``physical game'', ``movement game'', ``motion game'', ``motion-based game'', ``movement-based game'', ``training game'', ``sport(s) game'' or using similar terms (e.g., rowing game, table tennis game)\footnote{\rev{Without limiting the corpus to papers using the term exergame or variant words, we would have had to include almost every XR experience, given that even very simple interactions in spatial XR environments require some physical effort.}}.
    \item IC3: The exergame is an immersive XR exergame\footnote{\rev{Following the definition of immersive XR (see \autoref{sec:immersiveextendedreality})}}.
    \item IC4: The paper is written in English.
    \item IC5: The paper is not a theoretical paper (e.g., a literature review, a paper that does not include any self or commercial exergame implementation or its testing)\footnote{\rev{The paper does not have to include an evaluation (e.g., implementation-only papers are also included in our review)}.}.
    \item IC6: The paper is a research work (e.g., not proceedings preface).
    \item IC7: The paper does not have duplicates in the corpus.
\end{enumerate}

Using these inclusion criteria, two authors screened the corpus based on the titles and abstracts using the software \textit{Dovetail}\footnote{\url{https://dovetail.com/}}. Some papers, such as \cite{damasceno2022virtual}, had no abstracts, so we screened them based on their title and introduction. To screen the papers, two authors first reviewed the same $\sim$ 15\% ($N$=156) papers individually. Then, we conducted Cohen's kappa test~\cite{gamer2012package} to assess the inter-rater reliability between the two coders' binary decisions, which indicated \rev{almost} perfect agreement (\%94 of papers had agreed on, $\kappa$=0.872, $z$=10.9, \psmall); for the disagreements, one additional author acted as a tie-breaker in case the screening authors did not reach consensus after discussion, but this was never the case. Since we had \rev{almost} perfect agreement~\cite{mchugh2012interrater,landis1977measurement}, we split the remaining corpus between two coders, who screened the articles independently (one $N$=441, $N$=442). Such practices are common (e.g.,~\cite{rogers2021best}), and agreement strategies have been used in HCI literature to divide the data set between multiple coders~\cite{mcdonald2019reliability} to facilitate the coding process.

We also note that we excluded some articles during our screening phase that met our inclusion criteria because they did not use the required terms in the intended sense (e.g., training game for beekeepers~\cite{kellerhals2021let} and physical game to indicate that \rev{the} game has played in a physically shared environment~\cite{khoo2006age}). If cases were not clear, we included those papers to avoid missing any relevant work, such as work using the Kinect for interactions in VR (e.g., \cite{soares2019influence,hoda2015cloudbased,neves2015cardiovascular}) since this technology might be used as a tracking system with immersive XR.
After completing the screening, we kept 437 papers to be checked in the eligibility step.

\subsection{Eligibility}
For eligibility, we tried accessing all papers resulting from the screening ($n$=437). However, we could not retrieve two papers. Therefore, in this step, we reviewed a total of 435 papers based on their title, abstract, and full-text. In addition to the screening inclusion criteria, we also applied the following exclusion criteria (EC) for the eligibility check:

\begin{enumerate}[topsep=0pt, noitemsep, leftmargin=+.2in]
    \item EC8: Player(s) do not experience the exergame(s) through an immersive XR device.
    \item EC9: The paper uses commercial games without reporting the games' full titles.
    \item EC10: The paper presents an exergame but does not explain the game mechanics, i.e., which movements are performed.
    \end{enumerate}
    
Similar to the screening step, two authors coded $\sim$15\% of papers (n=66) independently. We used this step to check the inter-rater reliability between the two coders for the binary eligibility decision (agreement for 97\% of papers, $\kappa$= 0.939, $z$=7.64, \psmall). For any disagreements, we followed the same strategy from the screening step. Ultimately, we divided the remaining papers ($n$=371) between the same two authors (one $n$=185, $n$=186) because the inter-rater reliability showed \rev{almost perfect agreement}~\cite{mchugh2012interrater,landis1977measurement}. Based on our ICs and ECs, we excluded 249 articles.
We note that we observed some edge cases that resulted in exclusion: papers that include exercises in XR worlds but do not define the application as an exergame (e.g., \cite{granqvist2018exaggeration,li2022simulating}), papers that include a CAVE-like interface but do not offer a stereoscopic view (e.g., \cite{biffi2016immersive,martinniedecken2020hiit}), and papers that define their game as exergame but do not describe the included movements (e.g., \cite{wongutai2021effect}). Overall, this step resulted in 186 papers (see Appendix \autoref{tab:papers1} and \ref{tab:papers2}) being included for data extraction.

\subsection{Data Extraction}
We created a data extraction form using Airtable\footnote{\url{https://airtable.com/}} to extract the data from the included full-text papers. %
Initially, the lead author created a first version of the form, which was reviewed by the other co-authors. We then extracted the data for the first ten random papers and used the process to refine our form iteratively. Afterward, we applied the finalized data extraction form to all papers by dividing the remaining corpus between two coders ($n$=88, $n$=88). The final form had three parts (we supply the form and the extracted full data in the supplementary materials):
\begin{enumerate} [topsep=0pt, noitemsep, leftmargin=+.2in]

\item [(a)] \rev{\textit{General information.}}
First, we extracted general details of each paper \rev{(i.e., title, authors, publication year, and main objective)}.
\item [(b)] \rev{\textit{Study details.}}
Next, we assessed information on the conducted studies, focusing on the study design (i.e., type, independent and dependent variables) and recruited participants (i.e., number, sample details). If publications had multiple studies, we completed the form once for each. We only considered studies where participants actively played at least one version of the exergame and provided feedback. For example, we would complete the form separately for a multi-session HCD process and a subsequent independent user study of the final prototype (e.g., \cite{247}). In contrast, we excluded purely exploratory gameplay sessions of arbitrary VR games (i.e., not exergames) or domain expert interviews preceding the actual implementation phase (e.g., contextual inquiry~\cite{12}). We also did not extract information about experts' feedback, where those did not actively play the games but only reacted to video material (e.g., \cite{156}).
\item [(c)] \rev{\textit{Information about the used/developed exergame.}}
Lastly, we extracted details on the featured exergames, considering only commercial or custom exergames using immersive XR. Apart from general information \rev{(i.e., number, names, and duration of games)}, we mainly focused on GPEDT-related information. %
\end{enumerate}

Overall, this process led to a total of 200 filled forms (i.e., studies) for 186 papers, which can be seen in the supplementary materials.
During the data extraction, we encountered papers using commercial exergames. For these games, we performed an additional data extraction step and gathered information from the following resources: Steam\footnote{\url{https://store.steampowered.com/}}, Meta Store\footnote{\url{https://www.meta.com/experiences/}}, PlayStation Store\footnote{\url{https://store.playstation.com/}}, and the game publisher's website. We only used this official information because (i) authors typically include only relevant information for their study, and (ii) the publisher is the primary source for their game.

\subsection{Synthesis}
For our corpus, we used two types of analysis method: frequency-based reporting and taxonomy development methodology~\cite{nickerson2013method}.

\subsubsection{Frequency-based Analysis} Our goal with this analysis was to provide an overview of existing XR exergames research, covering published papers, conducted studies, and featured games. This information provides an understanding of the current state, a reference point for future studies, and can help identify research gaps.

\subsubsection{Taxonomy Development Methodology}

To create a hierarchical taxonomy of XR exergames applications, we followed the \rev{principles of the widely used taxonomy development methodology of}~\citet{nickerson2013method}, who define taxonomies as \quoting{systems of groupings that are derived conceptually or empirically}. A taxonomy should be \textit{useful}, which means it should be \textit{concise}, \textit{robust}, \textit{comprehensive}, \textit{extendible}, and \textit{explanatory}. \rev{\citet{nickerson2013method} recommends a set of objective ending conditions to create a taxonomy iteratively. From those, we used the first ending condition for our taxonomy (\quoting{All objects or a representative sample of objects have been examined}) and we analyzed all identified papers in our scoping review iteratively (see below for more details).}

To start the taxonomy development, researchers should decide on \textit{meta-characteristics}, which are \quoting{the most comprehensive characteristic[s] that will serve as the basis for the choice of characteristics in the taxonomy}~\cite{nickerson2013method}. Based on these meta-characteristics, researchers develop dimensions and characteristics within each dimension. Each dimension is \textit{mutually exclusive} (i.e., \quoting{no object can have two different characteristics in a dimension}~\cite{nickerson2013method}) and \textit{collectively exhaustive} (i.e., \quoting{each object must have one of the characteristics in a dimension}~\cite{nickerson2013method}). 
We note that we \rev{chose not to follow the mutual exclusivity and collective exhaustivity aspects (similar to other works~\cite{hertel2021taxonomy})}, as they may result in the loss of details (e.g., XR exergames might be designed for multiple user groups), and not every game description sufficiently reports the necessary details. 

\paragraph{Meta-characteristics \& GPEDT}
According to \citet{nickerson2013method}, the choice of meta-characteristics \quoting{should be based on the purpose of the taxonomy}. The objective of our taxonomy is to systematically analyze, organize, and map the current XR exergame landscape through the lens of HCI. Accordingly, we based our choice on the PACT framework and our prior domain knowledge.
\rev{We decided on our final meta-characteristics based on a discussion between four researchers, resulting in}~\groupgoals~Goals, \grouppeople~People, ~\groupactivities~Exercises,~\groupcontext~Design, and~\grouptech~Technologies (GPEDT). \groupgoals~Goals refers to the purpose of the designed/used XR exergame(s). \grouppeople~People considers any information about the target audience of the XR exergame(s).\groupactivities~Exercises focus on information about the exercises included in the XR exergame(s). \groupcontext~Design is concerned with any game design information relating to the XR exergame(s). Finally, \grouptech~Technologies captures information about the technology (hardware, devices) used in XR exergame(s).

\paragraph{Approach}
\rev{For the taxonomy development, the lead author first created deductive codes based on our meta-characteristics:} For \groupgoals~Goals, we consulted various resources to derive our deductive codes~\cite{kappen2019older,national2021four} (e.g., cognitive training).
For \grouppeople~People, we created deductive categories that capture the users' age groups (e.g., older adults) and, \rev{if applicable, their clinical conditions (e.g., dementia)}.
\groupactivities~Exercises covered three categories (e.g., body parts) containing different codes (e.g., full body). \groupcontext~Design focused on four overarching categories like task adaptation, each with their own codes (e.g., physiological measure-based adaptation). Finally, \grouptech~Technologies included three deductive categories (e.g., display) and corresponding codes (e.g., VR HMD).

\rev{For coding, we followed a hybrid approach. Our strategy involved the aforementioned deductive codes and data-driven inductive codes.}
First, two authors randomly selected ten exergames to code iteratively using the \rev{deductive categories and codes}. This step attempted to produce uniformly understood codes. After each paper, we reviewed our codebook understanding in a meeting. If necessary, we refined and created new inductive codes. \rev{For example, although we used deductive codes from \citet{kappen2019older} for the goals dimension (e.g., rehabilitation), we crafted new codes (e.g., preservation) for a more nuanced understanding of the aim and design of exergames.} After creating a shared understanding, we assigned the remaining dataset to be coded by the two coders independently. Both authors met multiple times to discuss challenging topics and newly arisen codes. They refined their codebook during the process and established coder consistency if needed. The full-text list of our codes is in our supplementary materials.

After coding, we created digital notes from our final codes and performed affinity mapping (see the supplementary materials) with the two researchers using a Miro Board\footnote{\url{https://miro.com/}}. After establishing an initial taxonomy, we discussed it with a third author with games research expertise to finalize the dimensions and characteristics.

\rev{Overall, we followed both \textit{conceptual-to-empirical}---employing preexisting knowledge to a set of data-- and \textit{empirical-to-conceptual}---creating a taxonomy based on a set of data--- approaches to create the taxonomy~\cite{nickerson2013method}. We consider our priori focus of GPEDT as conceptual-to-empirical since the GPEDT concept was created based on the PACT framework and the research team's prior domain knowledge to ensure a clear focus on the essential elements from an HCI lens. The remaining steps of taxonomy creation (e.g., characteristics) followed an empirical-to-conceptual approach; the final taxonomy, including subdimensions and characteristics, arose through the scoping review and iterative data-driven approach}.

We provide a qualitative summary to explain our taxonomy and characteristics. Here, we used the codes from our codebook to produce summaries. Further, we provide a positionality statement on the authors' background~\cite{braun2021can,newton2012no}. Even after inter-rater-reliability steps and unifying the codes between the coders, the involved researchers might still affect the interpretation of results. The two authors who conducted the analyses have worked on the intersection of XR and exergames technologies for several years; they published papers in the XR exergame field, studied different XR technologies, and played and implemented XR exergames. The other two involved authors, working on games and XR respectively, added their own perspectives to the discussion and analysis. We also list the potential implications of this bias in our limitations.

\section{Results of Scoping Review on Extended Reality Exergames} %

\begin{figure*}
    \centering
    \includegraphics[width= 0.75\textwidth]{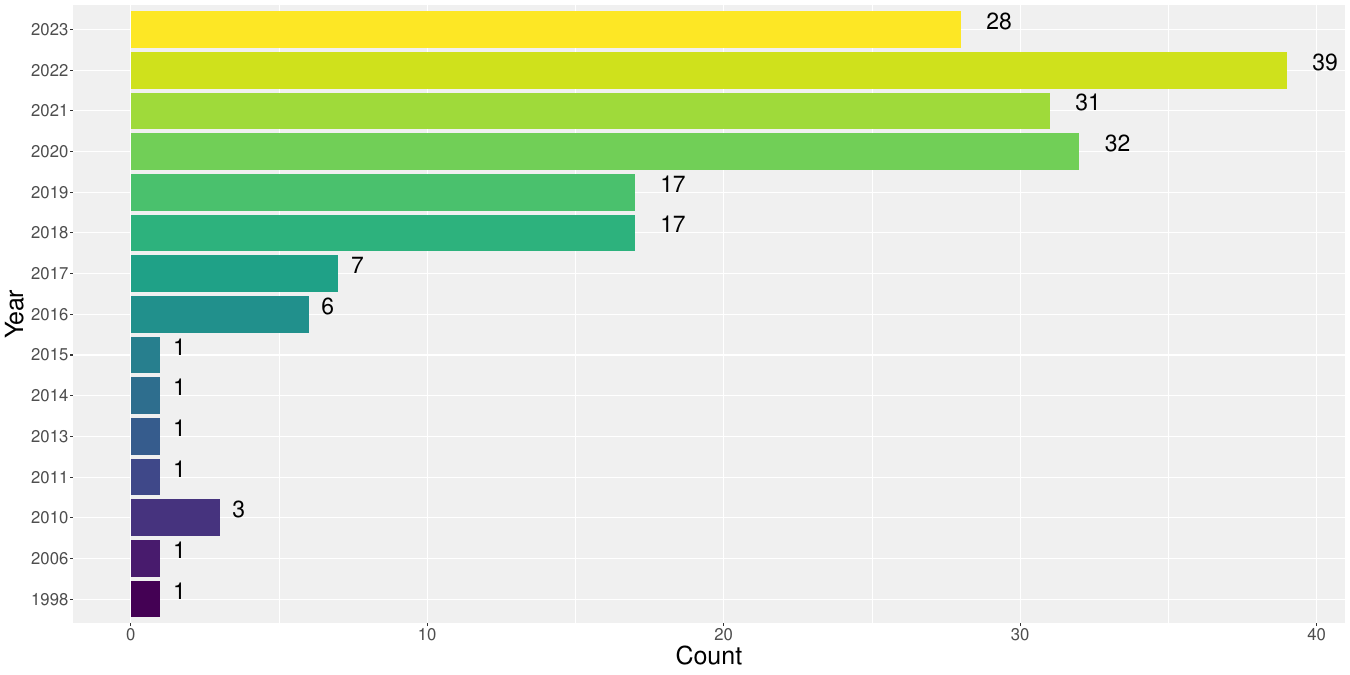}
    \caption{The distribution of our corpus ($N$=186) papers by year. We see a steep increase in published XR exergame papers.}
    \Description{The figure shows the distribution of our corpus ($N$=186) papers by year. The papers published between 1998 and 2023. We see a steep increase in published XR exergame papers.}
    \label{fig:year}
\end{figure*}

\begin{figure*}
    \centering
    \includegraphics[width= 0.8\textwidth]{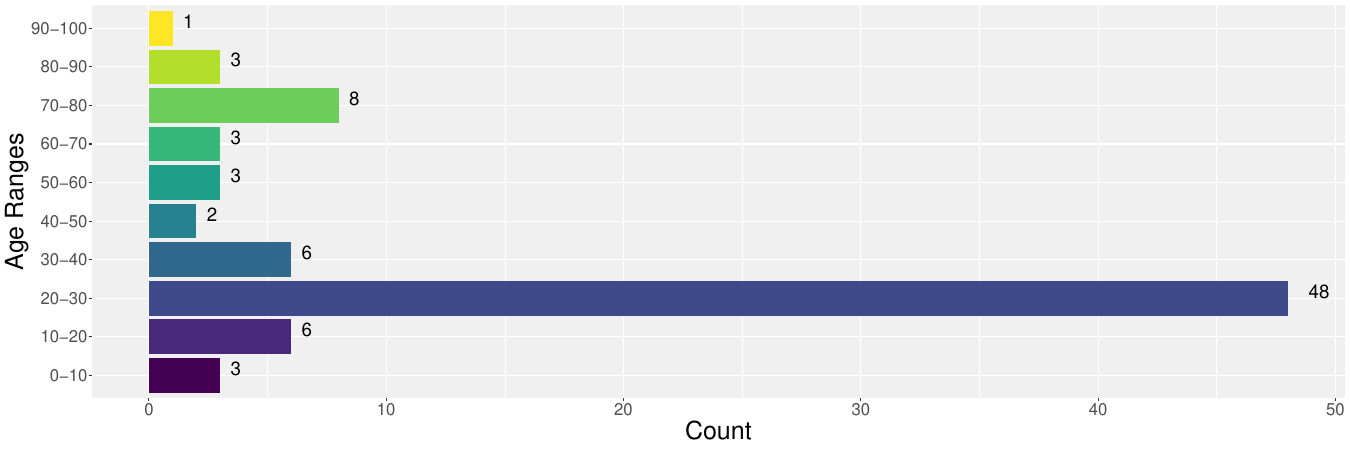}
    \caption{The average age distribution of our corpus' studies ($n$=83) by age range group. We see that most studies included participants with an average age range of 20-30 ($n$=48).}
    \Description{The figure shows the distribution of our corpus' studies ($n$=83) average age by age range group. We see that most studies included participants with an average age range of 20-30 ($n$=48).}
    \label{fig:age}
\end{figure*}

Here, we report the quantitative findings of our scoping review before explaining our hierarchical taxonomy in~\autoref{sec:tax}.
We included a total of 186 articles ranging from 1998 to 2023 in our corpus. The steep increase visible in \autoref{fig:year}---over 50\% of the papers were published in the last three years---shows that XR exergame research is booming and producing new publications rapidly. 

\subsection{XR Spectrum}
The majority of research focused on PC-tethered VR headsets ($n$=136). This group is dominated by the HTC Vive (including Pro, $n$=80), Oculus Rift ($n$=39), and Valve Index ($n$=5). Other headsets, such as the HP Reverb, Lenovo Explorer, or PlayStation VR, were only used in one paper each. Although much younger than PC-VR, mobile VR headsets take second place ($n$=42). Besides the popular Meta Quest platform (comprising Meta Quest 1 and 2, $n$=33), eight papers also used older 3DOF headsets like Google Cardboard, Oculus Go, or Samsung Gear VR. In contrast, projection-based VR systems, like CAVE, were only featured in seven publications. Lastly, nine papers explored AR exergames, utilizing Microsoft's HoloLens ($n$=6), the pass-through functionality of the HTC Vive Pro ($n$=1), or a custom-built solution ($n$=1).

\subsection{Study \& Sample Characteristics}

Thirty-three papers (17.74\%) presented only an exergame implementation. Of the remaining 153 papers, 9 conducted an HCD/iterative design approach with, on average, 15.5 participants (\sd10.27, one not reported). Four papers complemented the HCD with a final user study to test their product. Except for one publication, all HCD studies targeted older adults or people with dementia. Unfortunately, only four of these HCD papers reported sufficient demographic data: the 80 participants in these publications were \rev{mostly} older female adults with a mean age of 80.17 (\sd9.05, woman=63, man=17).

Apart from the HCD approaches, we extracted 158 evaluatory user studies. Although most papers only covered one user study, eight papers featured two or even three studies. The 158 studies included, on average, 24.63 participants (\sd26.50, three not reported) and covered a broad range from single-participant case studies to large evaluations reaching 250 players. 

Unfortunately, many papers miss crucial demographic information. Even after recovering the missing data to the best of our capabilities (i.e., performing age merge of different groups tested in the papers), only 83 studies (54.25\%) report both the mean and standard deviation of the sample population's age. Across these studies, the mean age was 31.67 (\sd19.76, $N$=2166, see~\autoref{fig:age}).

Similarly, gender distribution was only fully reported in 106 cases (67.09\%). The sample was slightly skewed, with 53.70\% men compared to 46.30\% women ($N$=2421, no non-binary or other). 

For 65 studies, papers reported the sample culture: most studies were conducted in Europe (UK: 12, Germany: 7, Norway: 4, Spain: 4, Finland: 2, Greece: 2) and North America (USA: 15, Canada: 1). Only nine studies were conducted in Asia (China: 3, Japan: 2, South Korea: 2, Taiwan: 1, Malaysia: 1) and Oceania (Australia: 7, New Zealand: 2). Africa is only represented by one study~\cite{997} that was run simultaneously in the USA and Nigeria.

\subsection{Study Design \& Data Collection}

Our corpus contains 158 user studies. In 79.11\% of the cases, players experienced the exergame only once ($n$=125). Only 33 studies featured repeated play sessions, ranging from two runs on consecutive days to long-term exercise programs. %
To understand what the included papers explored in their evaluations, we analyzed the independent variables. Fifty studies evaluated factors relating to \emph{time}, such as differences between pre- and post-test scores ($n$=29), two consecutive play sessions ($n$=5), longer intervention periods ($n$=9), or exercise durations ($n$=1).
Similarly, \emph{game design}-related studies ($n$=47) received much attention. A notable example is incorporating gameplay elements into physical exercise ($n$=18). Others explored the influence of avatars, music, haptic elements, or narratives.
Other common topics include comparisons between different \emph{platforms} ($n$=24, e.g., VR vs. non-VR), \emph{exercise-related factors} ($n$=18, e.g., changes in difficulty), or the use of \emph{non-player characters} ($n$=11, e.g., for competition). 
Surprisingly, only a few studies compared different \emph{age and user groups} ($n$=8). The remaining categories are similarly underrepresented: \emph{augmenting movements} in XR ($n$=8), personalizing the exergame experience ($n$=6), creating \emph{multiplayer} experiences ($n$=5, \rev{i.e., multiple users play together)}, or providing visualized \emph{feedback} ($n$=3).

To explore the outcomes of independent variables, the studies employed a variety of subjective and objective measures. Above all, 79.11\% of the studies included quantitative subjective metrics (e.g., questionnaires) to assess players' experience ($n$=125). The most commonly covered aspects are cybersickness (Simulator Sickness Questionnaire (SSQ)~\cite{kennedy1993simulator}, $n$=31), intrinsic motivation (Intrinsic Motivation Inventory (IMI)~\cite{ryan2006motivational}, $n$=23), perceived physical exertion (Borg Rating of Perceived Exertion Scale~\cite{borg1998borg}, $n$=21), game experience (Game Experience Questionnaire (GEQ)~\cite{johnson2018validation,ijsselsteijn2013game}, $n$=19), and usability (System Usability Scale (SUS)~\cite{brooke1995dirty}, $n$=19). To quantify the effects of exergame interventions, authors recorded gameplay and movement data ($n$=53) and assessed physiological measures ($n$=60, e.g., heart rate). Few studies also relied on standardized physiological/cognitive tests ($n$=16), such as the Stroop test ($n$=5). Lastly, qualitative data was collected for 58 studies, typically through interview sessions ($n$=31) or open-ended questions ($n$=24).

\definecolor{goals}{rgb}{1.0, 0.33, 0.64}

\begin{figure*}[!h]
{
\resizebox{\textwidth}{!}{
\centering
\small
\includegraphics{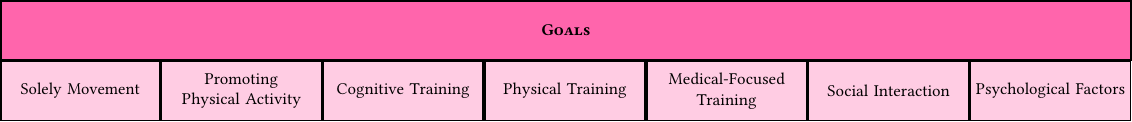}
}}
\caption{Our derived taxonomy's Goals dimension – with a total of seven characteristics.}\label{fig:goals}
\Description{The figure shows our taxonomy's goal dimension – with a total of seven characteristics.}

\end{figure*}

\subsection{Commercial vs. Custom XR exergames}

We faced three challenges in calculating the total number of featured XR exergames in extracted papers: (i) Some articles featured similar games, sometimes using the same name and similar visuals (i.e., we checked the figures). We considered these games to be unique because there were some adjustments to their design and implementation. (ii) Some articles included commercial XR exergames, which we considered only once. (iii) Some papers included systems/games that include multiple mini-games (e.g., Nvidia VR Fun house~\cite{nvidiafunhouse}, FitXR~\cite{fitxr}, \cite{412,338,230}); in these cases, we also treated these systems as one. Based on these criteria, we found 195 games: 20 commercial games (see \autoref{tab:commercial}) and 175 custom games.

\section{Taxonomy of Extended Reality Exergames}
\label{sec:tax}

We aimed to provide a taxonomy that guides the design, implementation, and research of future XR exergames. As our data unit, we consider all custom-built and commercial exergames as long as they were used in one of the papers. By analyzing the identified games according to the GPEDT aspects, we derived a taxonomy that is \textit{concise}, \textit{robust}, \textit{comprehensive}, \textit{extendible}, and \textit{explanatory}~\cite{nickerson2013method}. Since some dimension characteristics are not mutually exclusive, the reported numbers of games do not necessarily sum up. For example, the safety equipment characteristic comprises both stabilization harnesses ($n$=3) and monitoring devices ($n=7$). Since one specific game~\cite{338} requires both a harness and a monitoring screen, the broader safety category only contains nine distinct games.
In the following, we describe our taxonomy's dimensions and characteristics, and provide a qualitative summary of the corpus (see supplementary materials for the complete taxonomy illustration).

\definecolor{people}{HTML}{abdda4}

\begin{figure*}[!h]
{
\resizebox{\textwidth}{!}{
\centering
\includegraphics{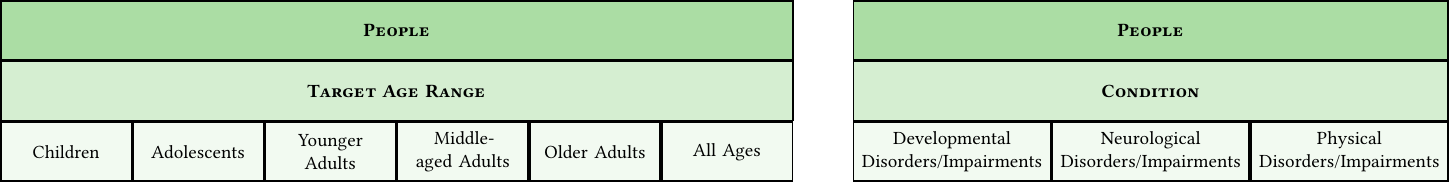}
}}
\caption{ (left) Our derived taxonomy's People subdimension: Target Age Range – with a total of six characteristics. (right) Our derived taxonomy's People subdimension: Condition – with a total of three characteristics.}\label{fig:agerange}
\Description{The left sub-figure shows our derived taxonomy's People subdimension: Target Age Range – with a total of six characteristics. The right sub-figure shows our derived taxonomy's People subdimension: Condition – with a total of three characteristics.}

\end{figure*}

\definecolor{activities}{HTML}{ffeda0}

\begin{figure*}[h]
{
\resizebox{\textwidth}{!}{
\centering
\small
\includegraphics{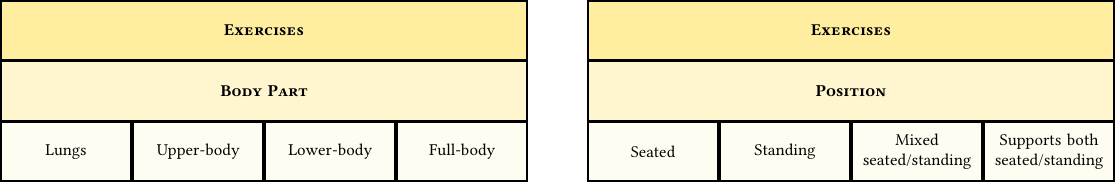}
}}
\caption{(left) Our derived taxonomy's Exercises subdimension: Body Part – with a total of four characteristics. (right) Our derived taxonomy's Exercises subdimension: Body Part – with a total of four characteristics.}\label{fig:bodypart}
\Description{The left sub-figure shows our derived taxonomy's Exercises subdimension: Body Part – with a total of four characteristics. The right sub-figure shows our derived taxonomy's Exercises subdimension: Body Part – with a total of four characteristics.}
\end{figure*}

\subsection{Dimension I: Goals}

The first dimension of our taxonomy is \groupgoals~Goals, explaining the inherent purpose of the XR exergames. For this dimension, we identified a set of seven distinct characteristics. Since many applications target multiple goals, this dimension is not mutually exclusive, i.e., XR exergames can have more than one goal. Our taxonomy's goal dimension is illustrated in \autoref{fig:goals}.

Overall, we identified 129 games that reported their purpose. 
Unsurprisingly, many XR exergames drive for \rev{physical training} ($n$=39); some games aim for endurance training ($n$=14) and involve high-intensity interval training protocols, such as \cite{460,160}. Others offer strength training for their users to improve muscular power ($n$=7). The remaining types of physical training targeted by XR exergames are balance ($n$=8, \cite{351,54}), skill ($n$=8, \cite{225,602}), flexibility ($n$=4, \cite{330,333}), and coordination/reaction training ($n$=7, \cite{171,291}).

\rev{Medical-focused} training is another essential characteristic of our goal dimension ($n$=52). A notable number of XR exergames ($n$=31) aim to provide rehabilitation opportunities for their users (e.g., \cite{338,syncsense,100}).
Other exergames ($n$=14) are designed as prevention/preservation tools; these games aim to prevent a deterioration of people's conditions and preserve their current state of abilities. Some games ($n$=8) are used to monitor/test the current abilities of users (e.g., range of motion~\cite{505} or reactive control~\cite{547}). Finally, XR exergames are also used to train for everyday activities ($n$=5). 

Despite being less common than the first two characteristics, \rev{social interaction} is still an important goal of many analyzed games ($n$=15), which aim to connect multiple players in the virtual world. The \rev{psychological factors} ($n$=10) characteristic comprises games that target the psyche by improving psychological wellbeing ($n$=3, e.g., \quoting{to face depression effect}~\cite{605}) or by offering relaxation ($n$=7, \quoting{control of breathing: relaxation/mindfullness training}~\cite{123}).
Furthermore, \rev{cognitive training} is targeted in 14 games, e.g., \cite{230,6}.

Another prevalent goal was to \rev{promote physical activity} ($n$=57). This characteristic represents exergames designed to motivate users to perform more physical activity, for example: \quoting{The main focus of the exergame is to motivate full body movements to promote exercise [...]}~\cite{333}. 
Interestingly, in one game, the aim was \quoting{having the purpose of no purpose}~\cite{350}; the authors argued that moving is fun on its own without needing ``another'' purpose. We feature this intrinsic motivation in the characteristic \rev{solely movement}.

\subsection{Dimension II: People}
\grouppeople~People are the core elements of interactive systems. In the design of XR exergames, we identified two dimensions that concern people.

\subsubsection{Subdimension: Target Age Range}
This subdimension represents the target age group. Only 51 exergames provide information about their target age group range. Again, this subdimension is not mutually exclusive, i.e., a game can target different age groups.

This subdimension consists of a total of six characteristics. Nine games are designed for \rev{children} (e.g., \cite{406,181}). At the same time, \rev{adolescents}~\cite{364} and \rev{young adults}~\cite{123} are only targeted by one XR game each. Although we have seen the term ``younger generations'' or ``young people'' in the descriptions of a few games~\cite{982, 173}, we did not code them for this category. It was unclear what exactly the younger generation refers to; is it children or younger adults? While two games~\cite{367,173} are particularly designed for \rev{middle-aged adults}, the bulk of age-targeting exergames focuses on \rev{older adults} ($n$=31). Lastly, eight games aim to be inclusive to \rev{all ages}. \autoref{fig:agerange} shows our target age range subdimension of our taxonomy.

\subsubsection{Subdimension: Condition}
This subdimension maps people's medical conditions \rev{(i.e., disabilities and impairments)} into characteristics (see \autoref{fig:agerange} (right)). As one game can be designed for people with different health conditions, one game could have more than one condition characteristic.

In our corpus, 138 games do not report their user groups' condition, which leads to only 57 remaining XR exergames falling into one of three categories.
First, the \rev{developmental disorders/im\-pair\-ments} characteristic ($n$=8) describes people who typically have difficulties with attention, learning, or using certain skills, such as language. In our corpus, XR exergames primarily target people with attention deficit hyperactivity disorder ($n$=4,~\cite{406,181}), autism ($n$=3, \cite{823,841,806}), and intellectual disabilities ($n$=1, \cite{720}). 

\rev{Neurological disorders/impairments} ($n$=26) covers all neurological conditions, such as people with dementia ($n$=11, \cite{46,279,301}), Parkinson's disease ($n$=6, \cite{115, 352, 1}), and ataxia ($n$=2, \cite{265,1036}).

The final characteristic focuses on \rev{physical disorders/im\-pair\-ments} ($n$=24), for example, upper limb conditions ($n$=4,~\cite{214,100,431}, hypertension ($n$=3, \cite{160,313}, and neck/back pain ($n$=2, \cite{80,127}).

\subsection{Dimension III: Exercises}
\groupactivities~Exercises dimension provides details on the movements performed in the XR exergames. We derived two subdimensions focusing on the targeted body part and supported play position to provide a comprehensive look at the movement aspect of exergames.

\subsubsection{Subdimension: Body Part}
This subdimension categorizes the XR exergames according to the body parts involved in the exercises (see \autoref{fig:bodypart} (left)). This subdimension is mutually exclusive, i.e., every game has one of the characteristics of the subdimension.

Most games ($n$=92) feature \rev{full-body} movements involving both the upper and lower body, e.g., by combining punches and squats. Secondly, 72 exergames focus only on \rev{upper-body} training, which we define as movements of the upper limbs, head, or torso: \quoting{After discussions with exercise therapists and considering safety, we decided to focus on upper body motions, including hand, shoulder, and head motions [...]}~\cite{906}. In contrast, fewer exergames ($n$=26) integrate solely \rev{lower-body} movements, i.e., using the lower extremities, into their gameplay: \quoting{A player moves through the scene by physically walking in the game area, a virtual narrow, winding platform over a virtual river of lava overlaid onto an empty hall or room}~\cite{991}. 
Finally, some exergames ($n$=5) also provide exercises involving the \rev{lungs}. For example, in \emph{Focus Tree}, \quoting{On inhaling, clouds cover the island to block the view of players; and on exhaling, clouds get blown away allowing players to view the island}~\cite{22}.

\definecolor{context}{HTML}{f46d43}

\begin{figure*}[!h]
{
\resizebox{\textwidth}{!}{
\centering
\small
\includegraphics{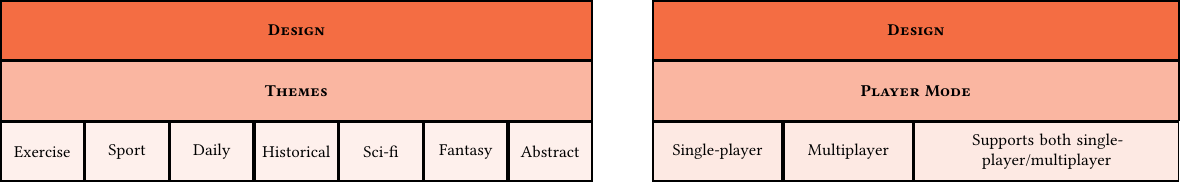}
}}
\caption{(left) Our derived taxonomy's Design subdimension: Themes – with a total of seven characteristics. (right) Our derived taxonomy's Design subdimension: Player Mode Setting – with a total of three characteristics.}\label{fig:theme}
\Description{The left sub-figure shows our derived taxonomy's Design subdimension: Themes – with a total of seven characteristics. The right sub-figure shows our derived taxonomy's Design subdimension: Player Mode Setting – with a total of three characteristics.}

\end{figure*}

\definecolor{context}{HTML}{f46d43}

\begin{figure*}[t]
{
\resizebox{\textwidth}{!}{
\centering
\small
\includegraphics{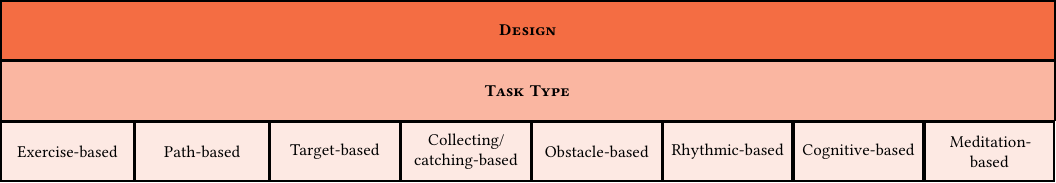}
}}
\caption{Our derived taxonomy's Design subdimension: Task type – with a total of eight characteristics.}\label{fig:tasktype}
\Description{The figure shows our derived taxonomy's Design subdimension: Task type – with a total of eight characteristics.}

\end{figure*}

\subsubsection{Subdimension: Position}
With this subdimension, we captured in which configuration the XR exergames are played: seated or standing. Determining this characteristic for every game was often challenging, since many games did not clearly state this information. Although many XR experiences could be playable in both configurations, we cannot simply guess without sufficient information on the implementation. Therefore, we decided to use the available sources in the following order: (i) review description, (ii) figures, (iii) supplementary materials, and (iv) available videos of the games. Based on this strategy, every game was coded under one characteristic (see \autoref{fig:bodypart} (right)). 

It was possible to retrieve the necessary information only for 187 games. Of these cases, 91 games (e.g., \cite{41}) feature exercises to be performed in a \rev{standing} position. A similar number of exergames ($n$=82) supports playing in a \rev{seated} position: e.g., \cite{1018,543,11}. Only two games (\cite{379,222}) included \rev{mixed} positions, i.e., players changed between those options to complete the entire game. Finally, 12 games \rev{supported both seated and standing} position gameplay: these games were mainly commercial games, e.g., ~\cite{beatsaber, fruitninja}.

\subsection{Dimension IV: Design}
This dimension, featuring four subdimensions, focuses specifically on \groupcontext~the game design.

\subsubsection{Subdimension: Theme}
The theme aims to describe the environment and inspiration of the visuals and gameplay of the exergame design. Our analyzed XR exergames fall into one of seven distinct characteristics to represent their themes (see \autoref{fig:theme} (left)).

The \rev{exercise} theme ($n$=19) features gym and exercise environments (e.g., gym hall) or solely focuses on exercise purposes; in \citet{379}'s game, players perform gestures in a mostly empty virtual world. 
Closely connected is the \rev{sports} theme ($n$=35), which imitates real-world sports, for example, rowing on a lake~\cite{530} or skiing~\cite{95}.
\rev{Daily} life activities ($n$=58) are also featured in many XR exergames, e.g., collecting apples from a tree~\cite{1} or blowing candles~\cite{412}.
In contrast, the \rev{historical} theme is rarely used ($n$=2); \citet{67}'s game presents a \quoting{small medieval village that Vikings have just plundered}.
Exergames with elements of space flight and future technologies ($n$=11) were categorized as \rev{Sci-fi}. In \emph{Astrojumper}~\cite{841}, the players avoid obstacles in a space environment.
As one of the most common themes ($n$=51), the \rev{fantasy} theme incorporates fantastic scenarios, non-real objects, and superpowers (e.g., \cite{131,566,65}). For instance, in the \emph{GhostStand} game, players beat ghosts~\cite{566}. %
\rev{Finally, the \rev{abstract} theme consists of games with a minimal design ($n$=19), such as \cite{beatsaber,456}; Beat Saber~\cite{beatsaber}'s game elements include stylized cubes and rectangles.}

\subsubsection{Subdimension: Player Mode}
This subdimension considers the player mode of XR exergames (see \autoref{fig:theme} (right)). This subdimension has three characteristics; each game is assigned to one. \rev{Single-player} covers XR exergames that are designed to be played only by one player at a time ($n$=171), e.g., \cite{10,337}.
\rev{Multiplayer games} ($n$=12) enable more than one player to play the game together (e.g., \cite{1018,156}). However, some XR exergames, particularly commercial ones, \rev{support both single-player and multiplayer} modes, such as FitXR~\cite{fitxr}, HoloFit~\cite{holofit}, and \citet{6}.

\definecolor{context}{HTML}{f46d43}

\begin{figure*}[!h]
{
\resizebox{\textwidth}{!}{
\centering
\includegraphics{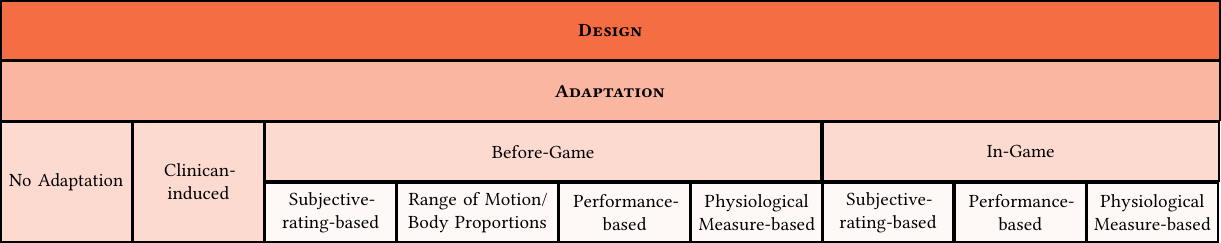}
}}
\caption{Our derived taxonomy's Design subdimension: Adaptation – with a total of four characteristics and seven sub-characteristics.}\label{fig:adaptation}
\Description{The figure shows our derived taxonomy's Design subdimension: Adaptation – with a total of four characteristics and seven sub-characteristics.}
\end{figure*}

\definecolor{tech}{HTML}{3288bd}

\begin{figure*}[t]
{
\resizebox{\textwidth}{!}{
\centering
\includegraphics{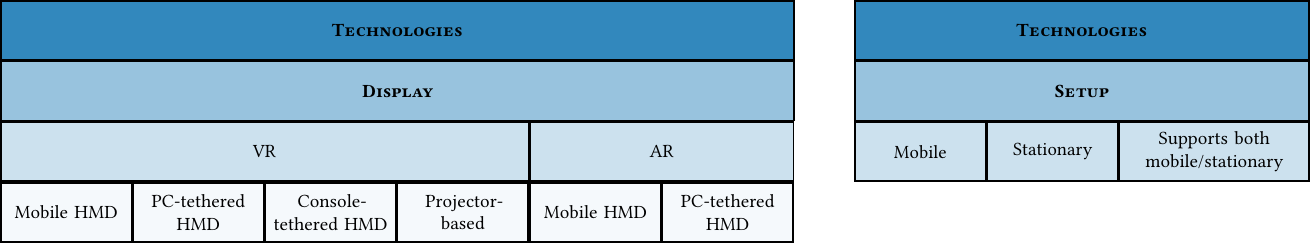}
}}
\caption{(left) Our derived taxonomy's Technologies subdimension: Display – with a total of two characteristics and six sub-characteristics. (right) Our derived taxonomy's Technologies subdimension: Setup – with a total of three characteristics.}\label{fig:display}
\Description{The left sub-figure shows our derived taxonomy's Technologies subdimension: Display – with a total of two characteristics and six sub-characteristics. The right sub-figure shows derived taxonomy's Technologies subdimension: Setup – with a total of three characteristics.}
\end{figure*}

\subsubsection{Subdimension: Task Type}

This subdimension presents the featured tasks in XR exergames. In total, we identified eight overarching characteristics. Since games are complex systems and typically include many tasks, one game can have multiple tasks.

\rev{Exercise-based} tasks ($n$=40) use standard workout exercises in a gamified environment. However, these movements are typically only partially embedded into the game's narrative. For example, players could perform squats that trigger magic attacks~\cite{139,367,131}.
With \rev{path-based} tasks ($n$=66), players travel along or follow a specific path in the game: In \cite{57}, \quoting{the player cycles along a straight path with a speed proportional to cycling revolutions per minute}.
In \rev{target-based} tasks ($n$=79), players must shoot, throw, or hit targets with their hands, weapons, or projectiles (e.g., \emph{VRabl}~\cite{982}).

Conversely, \rev{collecting/catching-based} tasks ($n$=45) feature approaching items that players catch or collect with their hands, weapons, or tools. While the previous two task types require players to interact with objects, \rev{obstacle-based} tasks ($n$=45) focus on avoiding objects. In \cite{333}, players must fit through holes in moving obstacles.
Next, \rev{rhythmic-based} tasks ($n$=10) include exercises and movements that players need to follow in rhythmic patterns (e.g., \cite{audiotrip,24}).
The XR exergames rarely featured \rev{cognitive-based} tasks ($n$=8). A notable example is \emph{Beat Saber}~\cite{beatsaber}, where players cut objects according to the direction shown on the objects.
Lastly, \rev{meditation-based} tasks feature tasks with a relaxation focus ($n$=5), for example, in the form of\quoting{halos [that] expanded and contracted when the player inhaled or exhaled, respectively, in real-time}~\cite{123}.

\subsubsection{Subdimension: Adaptation}

The adaptation dimension includes game task adaptations. A game may employ several adaption strategies. \autoref{fig:adaptation} illustrates this dimension.

The existing XR exergames mainly apply \rev{no adaptation} to their game tasks ($n$=125). We note that we do not consider calibration of the players' body proportions (e.g., height) if it is not framed as critical to the game's task since calibration should be done for every VR game to ensure that players can interact and that the avatar matches the players. Additionally, some games had adaptations that were controlled by clinicians or therapists (e.g., \cite{12,351}).

Another strategy was to apply \rev{game adaptation before gameplay} ($n$=45). Sixteen games adapt their task to the players' individual \rev{range of motion/body proportion}. For example, \cite{493} developed a game in which players pass through the holes in the walls and calibrated the cutouts to the players' body proportions to ensure comparable difficulties.
When \rev{subjective-rating-based} adaptation ($n$=18) is used, players can typically choose their preferred difficulty level (e.g., \cite{beatsaber,8}). Some games also employed \rev{performance-based} adaptation ($n$=7) before the gameplay, e.g., by adjusting the game based on the players' performance in a preceding task, such as~\cite{11,543}.
Lastly, few games used \rev{physiological-based} adaptation ($n$=5) before the gameplay; for example, \cite{14,11} adapted the resistance of their sports hardware based on players' body mass.

Besides adapting the task in an initial calibration phase, some games used \rev{in-game adaptation} strategies ($n$=25) during the gameplay. In-game \rev{subjective-rating-based} adaptation is only featured in one game ($n$=1) in the form of perceived exertion scores that contribute to adjusting the difficulty~\cite{458}.  
More commonly, in-game \rev{performance-based} adaptation ($n$=17) changes the game difficulty based on the players' success and performance (e.g., \cite{41, blackboxvr}.
Finally, in-game \rev{physiological-based} adaptation uses the players' physiological data to individualize the gameplay ($n$=9): for example, \quoting{if the heart rate is too low the procedurally generated rings are placed higher requiring more stroke power to pass through them}~\cite{460}.

\subsection{Dimension V: Technologies}
\grouptech~Technology is the last dimension of our taxonomy and provides information about the used technologies and their specifications.

\definecolor{tech}{HTML}{3288bd}

\begin{figure*}[!h]
{
\resizebox{\textwidth}{!}{
\centering
\small
\includegraphics{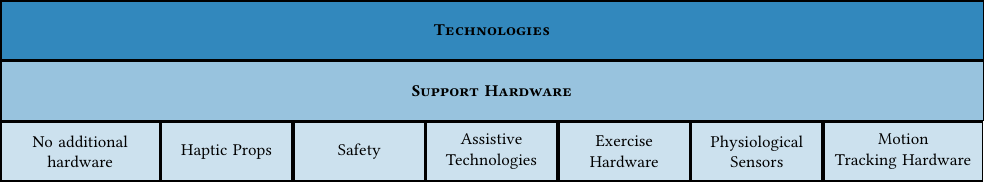}
}}
\caption{Our derived taxonomy's Technologies subdimension: Support Hardware – with a total of seven characteristics.}\label{fig:support}
\Description{The figure shows our derived taxonomy's Technologies subdimension: Support Hardware – with a total of seven characteristics.}
\end{figure*}

\subsubsection{Subdimension: Display}

The display subdimension categorizes which display technology is used by XR exergames (see \autoref{fig:display} (left)). Since one game can support more than one type of display, this subdimension is not mutually exclusive. For example, commercial games are often available for multiple platforms.

Most XR exergames use \rev{VR} displays ($n$=186). Many exergames rely on \rev{PC-tethered VR HMDs}, (e.g., HTC Vive or Oculus Rift), to display the games ($n$=141). Despite their recent popularity, \rev{mobile VR HMDs}, (i.e., standalone devices), only take the second place~\cite{1007,242,73} ($n$=48). Rarely, we also encountered the use of \rev{console-tethered VR HMDs} ($n$=6), (e.g., \cite{boxvr,beatsaber,carnivalgamesvr}). Apart from HMD-based VR, some games used \rev{projector-based VR} ($n$=7), (e.g., \cite{841,10,579}), which typically consists of projectors, walls, and 3D glasses.

The remaining XR exergames display their game world through \rev{AR} technologies, $n$=10. Seven of those games (e.g., \cite{982,100,1036}) used \rev{mobile AR HMDs}, such as Microsoft HoloLens~\cite{hololens}. The other three games relied on a PC and an attached \rev{PC-tethered AR HMD} to render the game, e.g., \cite{946,217,857}.

\subsubsection{Subdimension: Setup}
In this dimension, we give an overview of the supported setup conditions (see \autoref{fig:display} (right)). A single characteristic was allocated to each game.

The term \rev{mobile} ($n$=36) comprises games that work with portable hardware and thus can be played almost everywhere: \emph{Seas the Day}~\cite{279} only requires a mobile VR HMD, so players can run the game without being bound to one location.
\rev{Stationary} setups ($n$=149) cannot be easily moved to other places since they use hardware that is typically fixed to one location. Games in this category use headsets that are connected to a desktop PC~\cite{325}, bulky sports machines~\cite{554}, or permanently installed and calibrated motion-tracking hardware~\cite{338}.
Finally, ten commercial XR exergames \rev{support both mobile/stationary} setups, e.g., \cite{fruitninja,sportsscramble,dancecentral}.

\subsubsection{Subdimension: Support hardware}

The support hardware dimension comprises all additional hardware needed in addition to the regular display setup (e.g., VR HMD with controllers and base stations). Since games might require more than one type of supporting hardware, this dimension is not mutually exclusive.

Seventy-six games did not necessarily require \rev{any additional hardware} to be played (e.g., \cite{247,80,90}). In contrast, \rev{haptic props} ($n$=20) were used by some exergames. For example, \cite{1002} use a physical wakeboard for their exergame. Regarding \rev{safety} equipment ($n$=9), we encountered the use of \rev{harnesses} ($n$=3) and monitoring devices ($n$=7, e.g., displays~\cite{123,614}). In rare cases, XR exergames also included \rev{assistive} technologies ($n$=4), such as an exoskeleton ($n$=3, \cite{100,259}) or pneumatic gel muscles ($n$=1, \cite{95}).

\rev{Exercise hardware} ($n$=48) includes all balance, strength, and sports machines. Only two games use balance-exercising hardware: a robotic balance platform~\cite{128} and a balance board~\cite{222}. Only four games feature strength exercising hardware, which targets the strength development of muscles: weighted straps~\cite{214}, suspension bands~\cite{1002}, flywheel ergometer~\cite{916}, and cable resistance equipment~\cite{blackboxvr}.
Finally, sports hardware refers to equipment we use for training regardless of games, such as bicycle ($n$=30)~\cite{23}, elliptical ($n$=2)~\cite{syncsense}, rowing ($n$=9)~\cite{579}, and treadmill machines ($n$=4)~\cite{135}. 

\rev{Physiological sensors} ($n$=18) capture the hardware used to integrate physiological measures, such as heart rate ($n$=11)~\cite{662} and breath ($n$=8)~\cite{530}, in gameplay. In \cite{662}, heart rate was used to adjust the speed of a virtual competitor. In \citet{530}'s game, players were provided feedback based on their breathing rhythm.

\rev{Motion tracking hardware} is typically used in conjunction with the XR technology to track the players' body parts. In our corpus, 60 games used such hardware; the most often used were Vive Trackers ($n$=26, \cite{386,115,41}), followed by Kinect ($n$=20, \cite{283,709, 181}).

\section{Discussion}
This section reflects on previous XR exergame research, highlights future research directions, and provides recommendations for XR designers and developers. Second, we explain how to use our taxonomy: GPEDT. Finally, we provide guiding questions for systematic reporting of research on XR exergames.

\subsection{Reflecting on XR Exergame Research and Uncovering Research Directions}

\subsubsection*{\textbf{Continuum of Goals: XR Exergames are Enforcers?}}
The current XR exergame landscape has seven distinct goals. Almost all of them represent serious purposes; for example, promoting physical activity, providing medical-focused training, or physical training. But do XR exergames need to have serious goals? Should exergames force players to perform movements, or instead rely on players' inherent motivation to exercise~\cite{350}? Answering these questions is complex and requires serious discussion within the exergame research community. However, our taxonomy shows that the current decision of XR exergame design is on enforcing physical activity. Nevertheless, rare examples also show the promise of relying on players' inherent motivation. In \cite{40}, players were provided a custom grip controller to perform an optional strenuous activity that triggered virtual performance augmentation; as a result, players voluntarily performed strenuous activity for a longer period. \rev{Yet, we also note that our results are specific to the XR exergame research. Other movement-based applications (e.g., sports) may offer new means to interact with sports without enforcing players.}

\subsubsection*{\textbf{Different User Groups Require Different Tasks and Perhaps a Different Exergame.}}
Comparing the people dimension to our study data, \rev{only very few studies examined differences between user groups ($n$=8). This limited research attention is problematic since users with different abilities or physiologies play differently and prefer different tasks; \citet{12} reported that dementia-specific tasks bored older participants without dementia. Thus, further user group comparison studies are needed to inform the feature design of XR exergames to cater to user preferences and needs.}
Moreover, every XR application features at least some degree of movement because of its spatial interactions (e.g., grabbing). This raises the question of what to consider as exergames. Our taxonomy answers this question with ``it depends''. In our review, we followed the definition of \citet{muller2011designing}: \quoting{digital game[s] where the outcome [...] is predominantly determined by physical effort}. In our analysis, we saw that Nvidia Fun House~\cite{nvidiafunhouse} and Carnival Games VR~\cite{carnivalgamesvr} were considered as exergames~\cite{267,339} despite little physical activity. We believe that the physical effort required for a digital game depends highly on the user group, and researchers and developers should pay attention to this aspect.

\subsubsection*{\textbf{XR Exergames Knowledge is Limited When it Comes to Age Groups.}}
Our evaluation showed significant design and evaluation inconsistencies for the targeted age range. Most games with an age target focused on older adults. Also, the identified nine HCD approaches all targeted either older adults or people with dementia. On the other hand, most studies recruited a young audience: \rev{Only 24.10\% of studies had a sample with an average age above 40.} This discrepancy between the target user group and evaluation is problematic: Simply, the effectiveness of the exergame intervention and validity of measured player experience remains unclear. %
Furthermore, our results reveal that \rev{middle-aged adults~\cite{jochen2006adoption} are} rarely targeted by design or evaluation. Lastly, since abilities change drastically throughout life, we cannot assume that all exergames work for any age group. In contrast, we could even potentially harm people while driving for the good. In particular, the different physiology of children and adolescents requires special care. Hence, these age groups need more targeted exergame concepts.

\subsubsection*{\textbf{Generalizability of XR Exergame Research to Non-Western Countries.}}
When designing XR exergames, designers should consider people's cultural background because sociocultural factors influence how we use and accept technology~\cite{163}. Hence, the unbalanced culture sample identified in our corpus is worrisome. European and North American studies accounted for 72.31\%. The Global South, Asia, and Oceania are severely underrepresented. This raises doubts regarding the generalizability and relevance of XR exergames studies to non-Western cultures~\cite{linxen2021how}. Accordingly, additional research in non-Western nations is needed to understand better how sociocultural influences affect XR exergame design.

\subsubsection*{\textbf{\rev{Targeting Social Interaction and Multiplayer Engagement.}}}
Social interaction and multiplayer support received little attention. However, social settings are a critical factor in providing motivation to exercise. We assume this effect can apply to XR exergames, too. We suspect two primary explanations for the lack of coverage. First, most games feature single-owned HMD experiences because of limited hardware access. Here, we believe that asymmetric XR exergame experiences could be a promising solution~\cite{rogers2021best}. Second, real-life activities (e.g., video conversations) can satisfy social needs. Compared to these, XR technologies may not create the same social experience. Hence, people choose established options. Nonetheless, a recent work found that \quoting{VR (is) almost there}~\cite{sykownik2023vr} to mimic its real-world gaming counterpart. Therefore, we believe that future technology may improve XR exergame participants' connections. 

\rev{Further, we see potential in social interactions beyond traditional multiplayer setups (i.e., multiple users play together), such as the involvement of people who are not active players. For example, having non-player audiences that provide feedback~\cite{139}, or caregivers that guide the exergame experience~\cite{12} can contribute to the experience and performance of players. Although these types of social interaction have been covered in few publications~\cite{139,12}, or commercial games~\cite{fitxr}, we see the importance of further research as such interactions play a vital role in traditional training.}

\subsubsection*{\textbf{Designers and Developers Should Decide the Gameplay Position of Their Application at an Early Stage.}}
The games in our corpus are split almost equally between seated and standing gameplay. This spread is beneficial because both roles have crucial use cases. The standing setting can enable full-body exercises to combat our current lifestyle of sitting too much. However, many people may be unable to do this (e.g., because of age or space restrictions). %
Interestingly, research~\cite{379} showed that seated exergames increase physiological and perceived exertion. Contrary to common belief, seated exergames do not inevitably reduce physical activity. To be inclusive, it is best to support both seated and standing positions. However, only 6.42\% of our games supported this option. Of course, supporting both options might be difficult since most games must be designed from the ground up with this goal in mind. 
Hence, we advise developers to consider these early in their design process.

\subsubsection*{\textbf{\rev{Dominance of VR-HMD-based Exergames.}}}
\rev{In the past, XR exergames required stationary hardware, limiting their usability in the general public. New and powerful mobile headsets, such as Meta Quest or HoloLens, are a prominent step towards mobile and accessible exergames.} However, the drawback of most consumer hardware is limited motion tracking. Motion tracking of all four extremities is needed to create exergames that use the entire body. \rev{This is impossible without non-portable tracking devices like Vive Trackers or Kinect. But, Sony's Mocopi, HTCs inside-out trackers, and other future systems make us optimistic that this problem will be overcome quickly. Accordingly, many of the current stationary games might become mobile applications in the near future. However, other support hardware (e.g., rowing machines) still limits more advanced exergames to dedicated places.}

Our review focused on XR exergames, but VR dominated. Only 5.10\% of the analyzed games support AR HMDs. We presume the unavailability of AR HMDs is the primary cause of this lack of AR-focused research. The primary AR system used by games in our corpus is the Microsoft HoloLens. Although it advances mobile and lightweight AR glasses, the HoloLens has a restricted field of view, gesture-only interactions, and a substantial retail price. These shortcomings hinder AR development and acceptance, even though, for specific user groups, AR might be preferable over HMD-based VR systems. For example, people with neurological conditions like dementia likely profit from preserving a connection to the real world~\cite{12}. Similar problems also exist for CAVE-based systems. While they boost proprioception since users can see their bodies, the major financial and structural restrictions precluded wider use.

\subsubsection*{\textbf{Long-Term Effects and Adherence of XR Exergames Remain Unclear.}}
Exergames may promote physical activity and help prevent sedentary behavior globally. However, it is uncertain if games provide enough incentives for adherence needed for behavior change. Also, the long-term effects of XR exergames are poorly studied---only nine studies evaluated effects over more than two play sessions. Thus, the practicality of such games beyond gameplay novelty remains unknown. Hence, further research in two complementary directions is crucial: (i) conducting more long-term studies and (ii) exploring the raised knowledge gaps to create a good foundation for designing games with strong adherence.

\subsubsection*{\textbf{\rev{Transferability of Our Results to General XR \& HCI Research}}}
\rev{Our results represent the subsample of XR and HCI research, and align with previous papers examining a broader area in HCI. For example, \citet{linxen2021how} found that non-Western participants are underrepresented at CHI. Similarly, the novelty effects of XR reported by many researchers~\cite{karaosmanoglu2021feels,porter2019analysis,rogers2018vanishing} emphasize the need for long-term exploration of XR technologies in general.}

\rev{We see the potential for our results to be applied to broader HCI research beyond XR exergames. We believe that many of the guiding questions and taxonomy dimensions can be easily applied to other XR games or applications. Every XR research application has a purpose, there are people who will use the system, every XR application typically involves spatial interaction (even if limited), every application has a design that matches its research intention, and is used with certain technologies. However, we also emphasize that the more exergame-specific dimensions (e.g., exercise dimension) are not easily transferable. Lastly, given that convenience sampling is a common approach to recruit participants (e.g., among students), we speculate that bias in the representation of certain age groups may be applicable to broaden HCI research, but further research is needed to inform this.}

\subsection{How to Use the Taxonomy of XR Exergames}

With our taxonomy, our goal is to create a more standardized and systematic approach to XR exergame design, research, and communication in the field.
For example, exergames aimed at specific outcomes could be developed through a taxonomical lens presented within the GPEDT. With a building block approach, researchers and practitioners can explore how exergame features can effectively incentivize individuals, for example, to promote social interaction. Starting with the predetermined factors, they would explore the taxonomy iteratively to determine the best-fitting characteristic for every dimension. A populated example of such an exergame could ultimately look like this:
\begin{figure}[h]
\texttt{To improve \colorbox{goals!30}{[social interaction]} among\\ \colorbox{people!50}{[adolescents]}, use \colorbox{activities!30}{[full-body]} movements in a\\ \colorbox{activities!30}{[standing]} position to build a \colorbox{context!30}{[sci-fi]}-themed exergame featuring a\\ \colorbox{context!30}{[rhythmic-based]} \colorbox{context!30}{[multiplayer]} task using \colorbox{tech!30}{[mobile AR-HMDs]}}. \\
\end{figure}

In addition, our taxonomy can act as a gap analysis tool. Researchers can readily find underrepresented areas or untapped opportunities by mapping XR exergames based on the frequencies reported in each characteristic (see supplementary materials for each coded game); if the taxonomy reveals that there are few exergames designed for older adults, that is a clear area for future research. Similarly, if most existing games focus on physical training but not social interaction, that is another avenue for innovation.

\aptLtoX{\begin{table*}[!t]
\caption{The list of guiding questions to follow when reporting details of implemented/used XR exergames.}
\Description{The table lists the guiding questions that we provide for researchers, designers, developers to follow when they report details of implemented/used XR exergames.}
    \centering
    \small
    \resizebox{\textwidth}{!}{
    \begin{tabular}{ll}
    \toprule
    \multicolumn{1}{l}{\textsc{\textbf{No}}} & \textsc{\textbf{Guiding Questions}}\\
    \midrule
    \multicolumn{2}{l}{\groupgoals \xspace \textbf{Goals}} \\
    \cline{1-2}
    \rowcolor{goals!15} \#1 & What is the purpose of the XR exergame? \\
    \midrule
    \multicolumn{2}{l}{\grouppeople \xspace \textbf{People}} \\
    \cline{1-2}
    \rowcolor{people} 
    \#1 & Which age group does the XR exergame target? \\
    \rowcolor{people} \#2 & For which clinical group was the XR exergame designed or is it safe for everyone to play it? Why? \\
    \midrule
    \multicolumn{2}{l}{\groupactivities \xspace \textbf{Exercises}}\\
    \cline{1-2}
    \rowcolor{activities} 
    \#1 & Which movements do players perform in the XR exergame? \\
    \rowcolor{activities} \#2 & How do players precisely perform the featured movements in the XR exergame? \\
    \rowcolor{activities} \#3 & Which body parts are targeted by the XR exergame?\\
    \rowcolor{activities} \#4 & In which position can/should the XR exergame be played? Why? \\
    \midrule
    \multicolumn{2}{l}{\groupcontext \xspace  \textbf{Design}}\\
    \cline{1-2}
    \rowcolor{context} \#1 & Which game tasks does the XR exergame feature?\\
    \rowcolor{context} \#2 & Is there any adaptation applied in the game to match the game tasks to the players' game skill, physiological state, or condition\\ \rowcolor{context!15} & If so, in which form is this adaptation applied? \\
    \rowcolor{context} \#3 & How are the game tasks completed (e.g., performing a throwing action using the arms)? \\
    \rowcolor{context} \#4 & What is the design theme of the XR exergame? Why? \\
    \rowcolor{context} \#5 & How many players can play the game at once? \\
    \midrule 
    \multicolumn{2}{l}{\grouptech \xspace \textbf{Technologies}}\\
    \cline{1-2}
    \rowcolor{tech} \#1 & Which technology is used to display immersive XR environment? Why (e.g., advantages)? \\
    \rowcolor{tech} \#2 & Is any additonal hardware necesssary to play the XR exergame? \\
    \rowcolor{tech} \#3 & Which type of setup is used for the XR exergame? Mobile or stationary? \\
\bottomrule
    \end{tabular}}
    \label{tab:reportingguidelines}
\end{table*}}{\begin{table*}[!t]
\caption{The list of guiding questions to follow when reporting details of implemented/used XR exergames.}
\Description{The table lists the guiding questions that we provide for researchers, designers, developers to follow when they report details of implemented/used XR exergames.}
    \centering
    \small
    \resizebox{\textwidth}{!}{
    \begin{tabular}{ll}
    \toprule
    \multicolumn{1}{l}{\textsc{\textbf{No}}} & \textsc{\textbf{Guiding Questions}}\\
    \midrule
    \multicolumn{2}{l}{\groupgoals \xspace \textbf{Goals}} \\
    \cline{1-2}
    \rowcolor{goals!15} \#1 & What is the purpose of the XR exergame? \\
    \midrule
    \multicolumn{2}{l}{\grouppeople \xspace \textbf{People}} \\
    \cline{1-2}
    \rowcolor{people!15} 
    \#1 & Which age group does the XR exergame target? \\
    \rowcolor{people!15} \#2 & For which clinical group was the XR exergame designed or is it safe for everyone to play it? Why? \\
    \midrule
    \multicolumn{2}{l}{\groupactivities \xspace \textbf{Exercises}}\\
    \cline{1-2}
    \rowcolor{activities!15} 
    \#1 & Which movements do players perform in the XR exergame? \\
    \rowcolor{activities!15} \#2 & How do players precisely perform the featured movements in the XR exergame? \\
    \rowcolor{activities!15} \#3 & Which body parts are targeted by the XR exergame?\\
    \rowcolor{activities!15} \#4 & In which position can/should the XR exergame be played? Why? \\
    \midrule
    \multicolumn{2}{l}{\groupcontext \xspace  \textbf{Design}}\\
    \cline{1-2}
    \rowcolor{context!15} \#1 & Which game tasks does the XR exergame feature?\\
    \rowcolor{context!15} \#2 & Is there any adaptation applied in the game to match the game tasks to the players' game skill, physiological state, or condition\\ \cellcolor{context!15} & \cellcolor{context!15}If so, in which form is this adaptation applied? \\
    \rowcolor{context!15} \#3 & How are the game tasks completed (e.g., performing a throwing action using the arms)? \\
    \rowcolor{context!15} \#4 & What is the design theme of the XR exergame? Why? \\
    \rowcolor{context!15} \#5 & How many players can play the game at once? \\
    \midrule 
    \multicolumn{2}{l}{\grouptech \xspace \textbf{Technologies}}\\
    \cline{1-2}
    \rowcolor{tech!15} \#1 & Which technology is used to display immersive XR environment? Why (e.g., advantages)? \\
    \rowcolor{tech!15} \#2 & Is any additonal hardware necesssary to play the XR exergame? \\
    \rowcolor{tech!15} \#3 & Which type of setup is used for the XR exergame? Mobile or stationary? \\
\bottomrule
    \end{tabular}}
    \label{tab:reportingguidelines}
\end{table*}}

\subsection{Reporting Standards in XR Exergames Research}

To promote comprehensibility, reproducibility, and transferability, academia relies on transparent communication of methodology and results. Good reporting standards are crucial, especially in highly interdisciplinary areas or when conducting reviews to structure a domain. XR exergame research is a young and emerging field that no reporting consensus exists for every aspect yet. Unsurprisingly, individual explanation styles caused some challenges during our data collection and coding process. Next, we provide some examples of incomplete, inconclusive, or problematic reporting types:

\begin{enumerate}[topsep=0pt, noitemsep, leftmargin=+.2in]
    \item Many papers do not report their XR exergame's goals ($n$=66). Similarly, 144 of 195 games do not mention their target audience. But, we cannot use the same games for the same goal or expect every exergame to be playable by or be harmless to everyone.

    \item We created our taxonomy by finding common patterns. However, while coding the exergames' movement data, we could not distinguish a pattern. The rationale was the reporting level of the XR exergame movements (please refer to the affinity mapping activity in the supplementary material to see the movement codes). Some XR exergames descriptions explain the movements from a higher level; for example, players have to perform leg and arm movements, dancing movements, or dodging movements, but what these movements feature is unclear. 

    \item Similar to previous concerns, it was unclear if exergames were intended to be played seated or standing. We had to consult additional resources (e.g., figures, supplementary materials, videos) to track down the missing information.

    \item We also used additional resources when the retrieved data did not report the display used for gameplay (e.g., figures, supplementary materials, gameplay videos).
\end{enumerate}

The only way to fix diverging reporting is to develop a shared understanding and common reporting patterns. As a first step towards a more systematic approach, we close this paper with guiding questions for researchers, designers, and developers to answer and communicate (see \autoref{tab:reportingguidelines}). Similar to other efforts~\cite{gerling2022reflections}, we emphasize the value of providing audiovisual supporting materials---videos showing gameplay sequences to illustrate how the movements translate into the game or executable of developed games to provide clear communication about the featured gameplay.

\subsection{Limitations}
In this section, we address the limitations of our methodology as well as the limitations of the corpus.

\subsubsection{Limitations of the Methodology}

We consulted prior XR and exergames publications, and conducted multiple informal searches to decide on our search query, but like every review, we cannot claim completeness. Despite our best efforts, our query might not have covered all terms. \rev{For example, different fields (e.g., movement science vs. HCI) may use different terms to refer to exergames, such as \textit{``active video games''} and \textit{``active games''} (e.g.,~\cite{martinez2022active}). 
Additionally, our query might have missed articles that used only the name of a sport (e.g., virtual rowing) instead of an exergame-related term. Since it is impossible to account for all sports in a query, we urge researchers to use at least one of the exergame terms (e.g., movement games) to ensure that their research is included in relevant literature reviews.}
Nevertheless, we believe that with our included search terms \rev{(23 keywords) (see \autoref{tab:newquery})}, we give a comprehensive overview of the research and design of XR exergames.

Aligning with \citet{nickerson2013method}'s methodology, our taxonomy can be \rev{easily} expanded given the rapidly growing field. \rev{For example, while this article was under review, \citet{kontio2023feel}'s paper on non-standing locomotion techniques (potentially useful for VR exergames) has been published, providing opportunities for further extension of the position subdimension (e.g., lying). Accordingly, we emphasize that taxonomies are rarely static constructs but will grow over time to include new research directions.}

Lastly, we elaborate on the reflexivity further~\cite{newton2012no,braun2021can}. The authors of this paper work on the intersection of the XR and exergames and have published on those topics for several years. However, none of them has in-depth knowledge of movement science or physiology. Although we believe their expertise contributes to understanding the XR exergame landscape, we acknowledge that an in-depth focus on movement science is not in this review.

\subsubsection{Limitations of the Corpus}
HCI and medical areas use different terminologies. For example, in HCI literature, VR refers to immersive digital worlds, whereas several medical papers use VR to refer to non-immersive digital worlds (e.g., games that are played on a TV). Another example was the description of hardware: HCI typically uses the term Vive trackers to refer to HTC Vive motion tracking hardware. However, a medical paper referred to this equipment as \textit{``pucks''}~\cite{898}. 
While every research field has its own established terms, we believe in the importance of having shared terminologies. The absence of mutual understanding limits communication and information flow between these fields. We hope that our work will be a first step towards finding shared terminologies, thereby supporting the growth of this interdisciplinary field.

Finally, we saw that medical literature describes performed movements in detail (e.g., shoulder flexion, abduction), while for HCI literature, the explanation is typically limited to the general action (e.g., arm movements, throwing). Moreover, for some articles, it was not clear which specific movement (e.g., which arm movement~\cite{1033}) was used, how certain exercises were accomplished (e.g., squatting while sitting~\cite{898}), or which movements were performed for the specific action (e.g., dodging in the \emph{Fishing Master}~\cite{406}). When we did not see how these movements translated to the gameplay, we omitted them from our frequency reporting. Hence, we do not claim that we provide the complete list of movements performed in XR exergames but provide an approximation in our best capabilities.

\section{Conclusion}

Motivation is a central requirement for preserving one's engagement in physical activity. XR exergames can support this by offering immersive virtual worlds with enjoyable gameplay. The increased public attention on this timely topic, paired with recent advances in XR technology, has led to a surge of research focusing on XR exergames. The rapidly expanding field calls for a comprehensive organization to steer the community's efforts toward unexplored but promising topics. Therefore, we conducted a scoping review of the current state of XR exergame research.Based on analysis of 186 papers, we give a quantitative and qualitative summary of XR exergame research. Further, we provide a taxonomy with five central dimensions that map the design space of XR exergames. Finally, we conclude with \rev{nine} research directions and guiding questions to guide future research and reporting in the XR exergame field.

\begin{acks}
This work was supported by the European Union, the German Federal Ministry of Education and Research (BMBF), the German Research Foundation (DFG), and partly funded under the Excellence Strategy of the Federal Government and the Länder. Additionally, this project was supported by the SSHRC INSIGHT Grant (grant number: 435-2022-0476), NSERC Discovery Grant (grant number: RGPIN-2023-03705), CIHR AAL EXERGETIC operating grant (number: 02237-000), CFI John R. Evans Leaders Fund or CFI JELF (grant number: 41844), and the Provost's Program for Interdisciplinary Postdoctoral Scholars at the University of Waterloo.
\end{acks}
\balance
\bibliographystyle{ACM-Reference-Format}
\bibliography{ref.bib}

\appendix
\onecolumn
\section{Supplementary Tables}
\begin{table*}[htb] %
\small
    \centering
    \caption{The final search queries in the syntax of each database. The final search was conducted on 11 August 2023.}
    \Description{This table shows the final search queries in the syntax of each database. The final search was conducted on 11 August 2023.}
    \begin{tabular}{p{0.15\textwidth}p{0.8\textwidth}}
        \toprule
         Database & Query \\
         \midrule
         The ACM Guide to Computing Literature & \texttt{Title:(("immersive" OR "VR" OR "AR" OR "AV" OR "MR" OR "XR" OR "virtual realit*" OR "augmented realit*" OR "extended realit*" OR "mixed realit*" OR "augmented virtualit*" OR "virtual environment*") AND ("exergame*" OR "exercise game*" OR "physical game*" OR "movement game*" OR "motion game*" OR "motion-based game*" OR "movement-based game*" OR "training game*" OR "exertion game*" OR "sport game*" OR "sports game*")) OR Abstract:(("immersive" OR "VR" OR "AR" OR "AV" OR "MR" OR "XR" OR "virtual realit*" OR "augmented realit*" OR "extended realit*" OR "mixed realit*" OR "augmented virtualit*" OR "virtual environment*") AND ("exergame*" OR "exercise game*" OR "physical game*" OR "movement game*" OR "motion game*" OR "motion-based game*" OR "movement-based game*" OR "training game*" OR "exertion game*" OR "sport game*" OR "sports game*")) OR Keyword:(("immersive" OR "VR" OR "AR" OR "AV" OR "MR" OR "XR" OR "virtual realit*" OR "augmented realit*" OR "extended realit*" OR "mixed realit*" OR "augmented virtualit*" OR "virtual environment*") AND ("exergame*" OR "exercise game*" OR "physical game*" OR "movement game*" OR "motion game*" OR "motion-based game*" OR "movement-based game*" OR "training game*" OR "exertion game*" OR "sport game*" OR "sports game*"))} \\ 
         \midrule
         Scopus & \texttt{TITLE-ABS-KEY (("immersive" OR "VR" OR "AR" OR "AV" OR "MR" OR "XR" OR "virtual realit*" OR "augmented realit*" OR "extended realit*" OR "mixed realit*" OR "augmented virtualit*" OR "virtual environment*") AND ("exergame*" OR "exercise game*" OR "physical game*" OR "movement game*" OR "motion game*" OR "motion-based game*" OR "movement-based game*" OR "training game*" OR "exertion game*" OR "sport game*" OR "sports game*"))} \\
         \midrule
         PubMed & \texttt{(("immersive"[Title/Abstract] OR "VR"[Title/Abstract] OR "AR"[Title/Abstract] OR "AV"[Title/Abstract] OR "MR"[Title/Abstract] OR "XR"[Title/Abstract] OR "virtual realit*"[Title/Abstract] OR "augmented realit*"[Title/Abstract] OR "extended realit*"[Title/Abstract] OR "mixed realit*"[Title/Abstract] OR "augmented virtualit*"[Title/Abstract] OR "virtual environment*"[Title/Abstract]) AND ("exergame*"[Title/Abstract] OR "exercise game*"[Title/Abstract] OR "physical game*"[Title/Abstract] OR "movement game*"[Title/Abstract] OR "motion game*"[Title/Abstract] OR "motion-based game*"[Title/Abstract] OR "movement-based game*"[Title/Abstract] OR "training game*"[Title/Abstract] OR "sport game*"[Title/Abstract] OR "sports game*"[Title/Abstract] OR "exertion game*"[Title/Abstract]))} \\
         \bottomrule
    \end{tabular}
    \label{tab:newquery}
\end{table*}

\begin{table*}[!h]
\caption{The list of used commercial games in XR exergames research.}
\Description{The table lists the used commercial games in XR exergames research.}
\small
    \centering
    \resizebox{0.9\textwidth}{!}{
    \begin{tabular}{llllll}
    \toprule
   \textsc{\textbf{No}} & \textsc{\textbf{Name}} & \textsc{\textbf{Papers}} & \textsc{\textbf{No}} & \textsc{\textbf{Name}} & \textsc{\textbf{Papers}} \\ 
    \midrule
    \#1 & Audio Trip~\cite{audiotrip}    & \cite{110,120} & \#11 & HoloPoint~\cite{holopoint}               & \cite{900} \\
    \#2 & Beat Saber~\cite{beatsaber}    & \cite{104, 117, 141, 163, 236, 397,421,465,900} & \#12 & NVIDIA VR Fun House~\cite{nvidiafunhouse}     & \cite{339}  \\
    \#3 & Black Box VR~\cite{blackboxvr}   & \cite{184} & \#13 & QuiVr~\cite{quivr}                   & \cite{465}  \\
    \#4 & BoxVR~\cite{boxvr}             & \cite{210, 274, 339} & \#14 & Snow Games VR~\cite{snowgames}           & \cite{284}  \\
    \#5 & Carnival Games VR~\cite{carnivalgamesvr}         & \cite{267} & \#15 & Sports Scramble~\cite{sportsscramble}         & \cite{236,980}  \\
    \#6 & Dance Central~\cite{dancecentral}          & \cite{163} & \#16 & SyncSense~\cite{syncsense}               & \cite{130}  \\
    \#7 & First Steps~\cite{firststeps}            & \cite{236} & \#17 & Thrill of the Fight~\cite{thrillofthefight}     & \cite{102}  \\
    \#8 & FitXR~\cite{fitxr}                  & \cite{163,239,282,328} & \#18 & VirZoom~\cite{virzoom}                  & \cite{218,436}  \\
    \#9 & Fruit Ninja VR~\cite{fruitninja}         & \cite{384, 900} & \#19 & VRSports Challenge~\cite{vrsportschallenge}      & \cite{997}  \\
    \#10&  HoloFit~\cite{holofit}               & \cite{111} & \#20 & VZFit~\cite{vzfit}                   & \cite{164}  \\
\bottomrule
    \end{tabular}}
    \label{tab:commercial}
\end{table*}

\begin{table*}
\caption{The list of papers included in the taxonomy development (N=186).}
\small
    \resizebox{0.65\textwidth}{!}{
     \begin{tabular}{llll}
    
   \toprule
  \textsc{\textbf{No}} & \textsc{\textbf{Authors \& Papers}}  & \textsc{\textbf{No}} & \textsc{\textbf{Authors \& Papers}} \\ 
    \midrule
    \increment & \citet{1002} & \increment & \citet{338} \\
    \increment & \citet{228} & \increment & \citet{350} \\
    \increment & \citet{333} & \increment & \citet{351} \\
    \increment & \citet{41} & \increment & \citet{352} \\
    \increment & \citet{546} & \increment & \citet{364} \\
    \increment & \citet{624} & \increment & \citet{367} \\
    \increment & \citet{784} & \increment & \citet{368} \\
    \increment & \citet{906} & \increment & \citet{373} \\
    \increment & \citet{1} & \increment & \citet{379} \\
    \increment & \citet{10} & \increment & \citet{385} \\
    \increment & \citet{1007} & \increment & \citet{40} \\
    \increment & \citet{1018} & \increment & \citet{402} \\
    \increment & \citet{1036} & \increment & \citet{406} \\
    \increment & \citet{11} & \increment & \citet{412} \\
    \increment & \citet{115} & \increment & \citet{431} \\
    \increment & \citet{12} & \increment & \citet{476} \\
    \increment & \citet{123} & \increment & \citet{480} \\
    \increment & \citet{127} & \increment & \citet{493} \\
    \increment & \citet{128} & \increment & \citet{5} \\
    \increment & \citet{129} & \increment & \citet{50} \\
    \increment & \citet{13} & \increment & \citet{505} \\
    \increment & \citet{131} & \increment & \citet{51} \\
    \increment & \citet{135} & \increment & \citet{514} \\
    \increment & \citet{139} & \increment & \citet{530} \\
    \increment & \citet{14} & \increment & \citet{54} \\
    \increment & \citet{144} & \increment & \citet{543} \\
    \increment & \citet{156} & \increment & \citet{554} \\
    \increment & \citet{160} & \increment & \citet{560} \\
    \increment & \citet{167} & \increment & \citet{561} \\
    \increment & \citet{171} & \increment & \citet{566} \\
    \increment & \citet{173} & \increment & \citet{57} \\
    \increment & \citet{181} & \increment & \citet{570} \\
    \increment & \citet{183} & \increment & \citet{579} \\
    \increment & \citet{193} & \increment & \citet{60} \\
    \increment & \citet{20} & \increment & \citet{605} \\
    \increment & \citet{202} & \increment & \citet{614} \\
    \increment & \citet{206} & \increment & \citet{644} \\
    \increment & \citet{21} & \increment & \citet{65} \\
    \increment & \citet{214} & \increment & \citet{662} \\
    \increment & \citet{217} & \increment & \citet{67} \\
    \increment & \citet{22} & \increment & \citet{710} \\
    \increment & \citet{222} & \increment & \citet{720} \\
    \increment & \citet{225} & \increment & \citet{725} \\
    \increment & \citet{23} & \increment & \citet{73} \\
    \increment & \citet{230} & \increment & \citet{8} \\
    \increment & \citet{234} & \increment & \citet{80} \\
    \increment & \citet{24} & \increment & \citet{806} \\

    \bottomrule
    \end{tabular}}
\label{tab:papers1}
\end{table*}

\begin{table*}
\caption{\autoref{tab:papers1} continued: The list of papers included in the taxonomy development (N=186).}
    \resizebox{0.6\textwidth}{!}{
     \begin{tabular}{llll}

  \toprule
  \textsc{\textbf{No}} & \textsc{\textbf{Authors \& Papers}}  & \textsc{\textbf{No}} & \textsc{\textbf{Authors \& Papers}} \\ 
  \midrule

    \increment & \citet{242} & \increment & \citet{82} \\
    \increment & \citet{247} & \increment & \citet{823} \\
    \increment & \citet{25} & \increment & \citet{841} \\
    \increment & \citet{251} & \increment & \citet{857} \\
    \increment & \citet{259} & \increment & \citet{87} \\
    \increment & \citet{265} & \increment & \citet{898} \\
    \increment & \citet{279} & \increment & \citet{90} \\
    \increment & \citet{283} & \increment & \citet{110} \\  
    \increment & \citet{286} & \increment & \citet{916} \\
    \increment & \citet{291} & \increment & \citet{94} \\
    \increment & \citet{3} & \increment & \citet{946} \\
    \increment & \citet{301} & \increment & \citet{95} \\
    \increment & \citet{313} & \increment & \citet{951} \\
    \increment & \citet{325} & \increment & \citet{952} \\
    \increment & \citet{375} & \increment & \citet{982} \\
    \increment & \citet{376} & \increment & \citet{984} \\
    \increment & \citet{380} & \increment & \citet{991} \\
    \increment & \citet{386} & \increment & \citet{120} \\
    \increment & \citet{443} & \increment & \citet{104} \\
    \increment & \citet{456} & \increment & \citet{117} \\
    \increment & \citet{458} & \increment & \citet{141} \\
    \increment & \citet{46} & \increment & \citet{163} \\
    \increment & \citet{460} & \increment & \citet{236} \\
    \increment & \citet{462} & \increment & \citet{397} \\
    \increment & \citet{463} & \increment & \citet{421} \\
    \increment & \citet{517} & \increment & \citet{465} \\
    \increment & \citet{526} & \increment & \citet{900} \\
    \increment & \citet{53} & \increment & \citet{184} \\
    \increment & \citet{545} & \increment & \citet{210} \\
    \increment & \citet{547} & \increment & \citet{274} \\
    \increment & \citet{55} & \increment & \citet{339} \\
    \increment & \citet{569} & \increment & \citet{267} \\
    \increment & \citet{58} & \increment & \citet{163} \\
    \increment & \citet{598} & \increment & \citet{239} \\
    \increment & \citet{6} & \increment & \citet{282} \\
    \increment & \citet{602} & \increment & \citet{328} \\
    \increment & \citet{709} & \increment & \citet{384} \\
    \increment & \citet{78} & \increment & \citet{284} \\
    \increment & \citet{100} & \increment & \citet{980} \\
    \increment & \citet{1021} & \increment & \citet{130} \\
    \increment & \citet{1028} & \increment & \citet{102} \\
    \increment & \citet{1033} & \increment & \citet{218} \\
    \increment & \citet{28} & \increment & \citet{436} \\
    \increment & \citet{324} & \increment & \citet{997} \\
    \increment & \citet{330} & \increment & \citet{164} \\
    \increment & \citet{337} & \increment & \citet{111} \\

\bottomrule 
    \end{tabular}}
\label{tab:papers2}
\end{table*}

\end{document}